\documentclass[10pt,letterpaper]{article}
\usepackage[margin=1in]{geometry}

\usepackage[T1]{fontenc}
\usepackage[english]{babel}
\usepackage[utf8]{inputenc}
\usepackage[normalem]{ulem}
\usepackage{booktabs}

\usepackage[round, semicolon]{natbib}
\usepackage{titlesec}

\usepackage{algorithm}
\usepackage{algpseudocode}

\usepackage{booktabs,multirow,tabularx,makecell}

\usepackage{fancyhdr}
\usepackage{amsmath}
\usepackage{upgreek}
\usepackage{siunitx}

\usepackage{amsmath,amsfonts,bm}

\newcommand{\J}{\mathbf{J}}

\def\eqref#1{equation~\ref{#1}}

\def\1{\bm{1}}

\newcommand{\rv}[1]{\bm{#1}}

\def\rvalpha{{\bm{\alpha}}}

\def\rvtheta{{\bm{\theta}}}

\def\rvpsi{{\bm{\psi}}}
\def\rvtau{{\bm{\tau}}}

\def\rvkappa{{\bm{\kappa}}}
\def\rvphi{{\bm{\phi}}}
\def\rvmu{{\bm{\mu}}}
\def\rvnu{{\bm{\nu}}}

\def\rvdelta{{\bm{\delta}}}

\def\rvb{{\mathbf{b}}}
\def\rvc{{\mathbf{c}}}

\def\rvx{{\mathbf{x}}}

\DeclareMathAlphabet{\mathsfit}{\encodingdefault}{\sfdefault}{m}{sl}
\SetMathAlphabet{\mathsfit}{bold}{\encodingdefault}{\sfdefault}{bx}{n}

\def\gF{{\mathcal{F}}}

\def\gN{{\mathcal{N}}}

\def\gU{{\mathcal{U}}}

\def\sR{{\mathbb{R}}}

\usepackage{amsthm}

\theoremstyle{definition}

\usepackage{amssymb,amsfonts}
\usepackage{textcomp}
\usepackage{gensymb}
\usepackage{tcolorbox}

\usepackage{graphicx}
\usepackage{float}
\usepackage{wrapfig}
\usepackage{caption}
\usepackage{subcaption}
\usepackage{multirow}

\usepackage{comment}
\usepackage{lmodern}
\usepackage[onehalfspacing]{setspace}

\usepackage{xcolor}
\definecolor{bahamablue}{RGB}{0, 104, 150}
\definecolor{apple}{RGB}{0, 104, 150} 
\definecolor{vinrouge}{RGB}{153, 76, 102} 
\definecolor{driftwood}{RGB}{178, 128, 76} 
\definecolor{brat}{RGB}{138, 206, 0}
\definecolor{atomictangerine}{RGB}{255, 177, 133}
\definecolor{amaranthpink}{RGB}{250, 158, 193}

\definecolor{codeborder}{RGB}{218, 218, 218}  
\definecolor{codebg}{RGB}{246, 246, 246}
\definecolor{codefont}{RGB}{177, 43, 70}    

\definecolor{brainyellow}{RGB}{194, 138, 75}
\definecolor{brainblue}{RGB}{68, 119, 170}
\definecolor{brainpurple}{RGB}{124, 107, 166}
\definecolor{braingreen}{RGB}{106, 155, 131}
\definecolor{secgreen}{RGB}{27, 158, 119}
\definecolor{simulatorgreen}{RGB}{27, 158, 119}

\newtcbox{\inlinecode}{on line, boxrule=0.5mm, colframe=codeborder, colback=codebg, rounded corners, boxsep=0mm, left=0.5mm, right=0.5mm, top=0.5mm, bottom=0.5mm}

\usepackage[colorlinks=true, urlcolor=apple, linkcolor=vinrouge, citecolor=bahamablue]{hyperref}

\newcommand{\review}[1]{\textcolor{black}{#1}}

\title{A High-Fidelity 3D Simulator for Synthetic fNIRS Data Generation}
\author{
    Condell Eastmond\thanks{Equal contribution.}\\
    \texttt{eastmc2@rpi.edu}
    \and
    Niels Bracher\footnotemark[1]\\
    \texttt{brachn@rpi.edu}
    \and
    Xavier Intes\\
    \texttt{intesx@rpi.edu}
    \and
    Stefan T. Radev\\
    \texttt{radevs@rpi.edu}
}

\begin{document}
\maketitle

\renewcommand{\sectionautorefname}{Section}
\renewcommand{\subsectionautorefname}{Section}
\renewcommand{\subsubsectionautorefname}{Section}

\begin{abstract}
\noindent\review{Functional near-infrared spectroscopy (fNIRS) provides a noninvasive window into brain activity by measuring task-related changes in oxygenated and deoxygenated hemoglobin in the cortex.
However, analyzing fNIRS data is challenging because of motion artifacts, physiological noise, and other confounding factors.
This challenge is further compounded by the limited availability of annotated datasets, which hinders the development and validation of new analysis pipelines, particularly given the growing use of AI methods.
Recognizing these challenges, we introduce a 3D fNIRS simulator that uses mesh-based Monte Carlo simulations to generate physiologically informed, full-head synthetic recordings with controllable spatiotemporal characteristics.
Our simulator combines anatomically informed sensitivity profiles with parameterized models of hemodynamic responses, systemic physiology, and measurement artifacts.
As a result, users can generate large labeled datasets with known cortical ground truth for developing and evaluating denoising algorithms, data augmentation strategies, mechanistic models, and \textit{in silico} experiments.
We validate key spatiotemporal and physiological properties of the simulator against experimental fNIRS data from open-source finger-tapping, pain-assessment, and surgical-skill datasets.}
\end{abstract}

\section{Introduction}
Significant innovations in neuroimaging now enable us to visualize the brain in action \citep{mcdowell2013real}. In particular, functional near-infrared spectroscopy (fNIRS) utilizes near-infrared light to detect changes in cortical hemodynamic activity. Driven by advances in system miniaturization, wearability, and system sensitivity \citep{doherty2023interdisciplinary}, fNIRS is uniquely able to monitor and quantify fast, spatially-resolved brain activations of mobile participants in naturalistic environments over long periods of time \citep{gao2024advances,pinti2020present,pinti2018review}.
Consequently, fNIRS has proven highly useful across various applications, including brain-computer interfaces \citep{naseer2015fnirs}, monitoring of brain disorders \citep{rahman2020narrative}, and neurofeedback \citep{klein2024lab}.

Despite its versatility, analyzing fNIRS data remains challenging. The primary goal is to extract cortical activation profiles from optical time series measurements that often contain features generated by numerous confounding factors, including motion artifacts, superficial signal contamination, and systemic physiological signals. These sources of noise obscure neural activity and complicate the accurate interpretation of measured hemodynamic responses. As a result, a wide range of refined analysis pipelines has been proposed; however, the field still lacks standardization and a universally accepted best practice \citep{yucel2025fnirs}. This fragmentation has also contributed to a scarcity of large, annotated datasets for cross-validation and for supporting the development of next-generation analysis pipelines, particularly those relying on AI \citep{doherty2023interdisciplinary,eastmond2022deep}. 

Herein, we argue that a high-fidelity fNIRS simulator can address these issues by providing access to flexible and controllable ground-truth data across different experimental configurations. Such a simulator would enable systematic benchmarking of signal processing algorithms using realistic ground-truths, while also providing a data augmentation pipeline for data-hungry AI models. In addition, it could enable simulation-based inference approaches \citep{cranmer2020frontier} that link neural time series to mechanistic models of subject-specific brain activity \citep{tolley2024methods}.

There are two general approaches to generating synthetic fNIRS data, each with distinct advantages and limitations. Recent advances in deep learning have enabled \textit{data-driven methods}, which train generative neural networks \citep[e.g., Generative Adversarial Networks; GANs,][]{NIPS2014_f033ed80} to learn the statistical structure of observed data and generate realistic synthetic samples \citep{rasheed2021generative, nagasawa2020fnirs, wickramaratne2021conditional}. For example, \cite{nagasawa2020fnirs} used a GAN to augment the training data of a neural classifier with synthetic fNIRS time series, resulting in an accuracy gain of around $30\%$. Generative AI approaches can capture complex patterns directly from data but often require large training annotated datasets and may lack interpretability or physiological grounding.

On the other hand, \textit{theory-driven} (or first-principles) data generators rely on biophysically detailed forward models whose parameters have a direct mechanistic interpretation \citep{tolley2024methods}. These theory-driven approaches have been used since the inception of fNIRS and have substantially benefited from increasingly accurate models of light propagation in complex biological tissues. For example, \citet{liu2022deep} employed a layered slab model of head tissue to generate synthetic fNIRS data for validating a deep learning-based filtering method. However, this approach was limited by its simplified anatomical representation. An earlier approach used a full head mesh to model functional connectivity, but did not attempt to simulate confounding factors such as systemic physiological contributions to fNIRS signals \citep{tak2015dynamic}. Moreover, many theory-driven models struggle to accurately capture spatiotemporal dynamics across the complex geometry of the human head \citep{tran2020improving}, or do not attempt to do so at all, and instead rely on simplified single-channel simulations \citep{leamy2011functional}.

\review{Hybrid and statistical approaches offer alternatives to fully synthetic fNIRS generation. For example, Cedalion combines spatially localized cortical responses, projected into channel space using an optical forward model seeded with experimental resting-state recordings \citep{middell2026cedalion}. While it preserves realistic background fluctuations and diversity, the pool of data it can generate remains constrained by the available recordings. Alternatively, MNE-NIRS can generate channel-level hemodynamic responses with autoregressive (AR) noise \citep{luke2021analysis,santosa2018nirs}. However, the statistical noise model does not explicitly distinguish physiological processes from measurement artifacts, while the lack of an anatomy-informed light-transport model limits the representation of spatial dependencies and prevents direct access to cortical ground truth. These limitations motivate a fully synthetic simulator for cortical activations with explicitly modeled physiological and unsystematic components and anatomically informed spatial sensitivity.}

\review{In this work, we present a fully synthetic 3D fNIRS simulator that combines anatomically informed light transport with parameterized models of cortical hemodynamics, systemic physiology, and measurement artifacts. Using mesh-based Monte Carlo (MMC) light transport \citep{fang2010mesh}, we generate spatial sensitivity profiles that map changes in optical absorption to channel-specific measurements according to head anatomy and optode geometry \citep{yao2018direct}. Cortical activity is modeled using region-specific hemodynamic response functions (HRFs), while systemic physiology includes cardiac and respiratory activity and Mayer waves \citep{hocke2018automated}. Additional measurement artifacts, including spikes, baseline shifts, and channel dropout, are modeled separately. Together, these components generate spatially structured fNIRS recordings with independently controllable signal contributions and known ground truth, while remaining mechanistically interpretable.}

\begin{figure*}[t]
    \centering
    \includegraphics[width=1.0\textwidth]{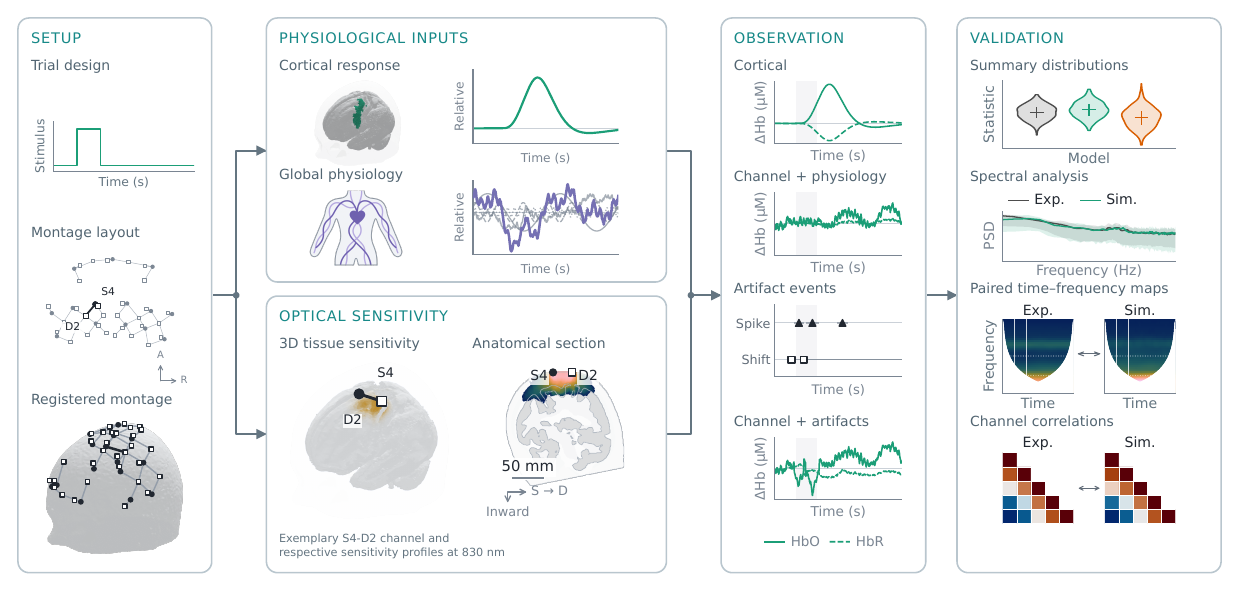}
    \caption{Graphical overview for simulating multichannel fNIRS data. Starting from an experimental design and optode montage (\textcolor{secgreen}{setup}), the simulator combines task-evoked cortical responses and global physiological fluctuations (\textcolor{secgreen}{physiological inputs}). A spatial sensitivity matrix derived from light-transport simulations (\textcolor{secgreen}{optical sensitivity}) translates these mesh-level components accurately into optode-level changes, producing synthetic long- and short-channel fNIRS time series with or without external artifacts (\textcolor{secgreen}{observation}). The simulated data can be compared with empirical recordings using global and local metrics, such as time-frequency maps, time-averaged spectral profiles, or inter-channel correlation maps (\textcolor{secgreen}{validation}).
}
    \label{fig:Pull_figure}
\end{figure*}

Starting with the theoretical foundations on light transport, we describe the mathematical foundations of our simulator and detail its parameter configurations. We then validate both low- and high-level spatiotemporal signal characteristics of the simulated data against experimental fNIRS measurements from publicly annotated datasets. 
\autoref{fig:Pull_figure} provides an overview of the proposed framework.

\section{Methods}\label{sec:Model}

The following sections introduce our multi-component forward model for generating synthetic fNIRS time series and establish the relationship between the forward model's parameters, denoted collectively as $\rvtheta \in \Theta$ (for an overview see \autoref{tab:parameters-overview}), and the time-dependent vector of intensities, denoted as $\rvphi_1(t; \rvtheta) \in \mathbb{R}^{C}$, where $C$ is the number of measurement channels (see \autoref{tab:dim-abbrev} for an overview of dimensional notation).

To establish the relationship between changes in tissue absorption and measurable light intensities (see \autoref{sec:observations}), we start by introducing the Rytov approximation \citep{oleary1996imaging}, which uses the baseline intensity $\rvphi_0$ and sensitivity matrix $\J$, both computed using MMC simulations (see \autoref{sec:sensitivity-matrix}).

\begin{figure}
    \centering
    \includegraphics[width=1.0\textwidth]{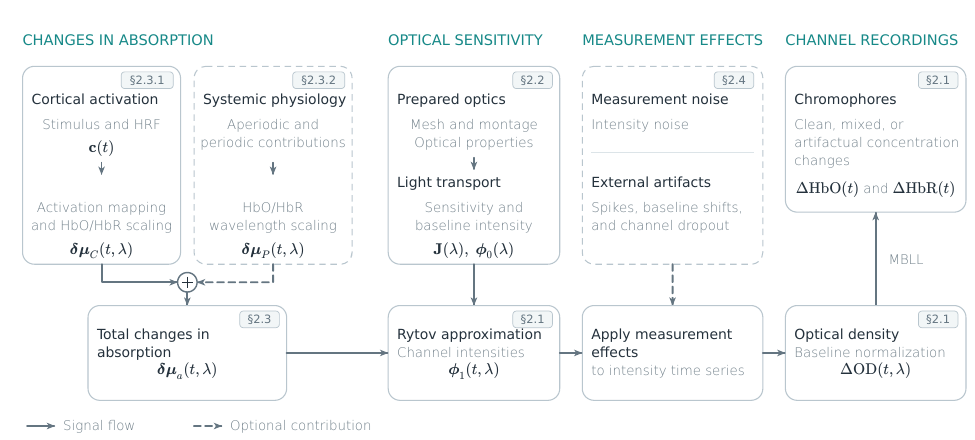}
    \caption{\review{Forward simulation paths and key components of the proposed fNIRS model with referenced method sections.}}
    \label{fig:model-overview}
\end{figure}

We then introduce the parameterization of cortical activity and systemic physiology used to generate spatiotemporal changes in tissue absorption, $\rvdelta\boldsymbol{\mu}_a(t;\rvtheta)$ (see \autoref{sec:Model}).
Lastly, we introduce a variety of external artifacts to more realistically model fNIRS data (see \autoref{sec:artifacts}). All model parameters and distributions introduced in the following are detailed in \autoref{sec:appdx-simulator-details}. 
\autoref {fig:model-overview} shows the resulting forward simulation paths of all model components combined.

\begin{table}[!ht]
    \centering
    \small
    \renewcommand{\arraystretch}{1.35}
    \setlength{\tabcolsep}{5pt}
    \begin{tabularx}{\textwidth}{lll>{\raggedright\arraybackslash}X}
        \toprule
        \multicolumn{2}{l}{\textbf{Parameter symbol}} & \textbf{Meaning} & \textbf{Description} \\
        \midrule
        \multirow{7}{*}{$\rvtheta_C$}
        & $\rv{d^+}=(d^+_1,\dots,d^+_L)$ & Positive gamma dispersions & Scales of the positive gamma components of regional HRFs. \\
        & $\rv{d^-}=(d^-_1,\dots,d^-_L)$ & Negative gamma dispersions & Scales of the negative gamma components of regional HRFs. \\
        & $\rvtau=(\tau_1,\dots,\tau_L)$ & Positive gamma modes & Times from stimulus onset to the modes of the positive gamma components. \\
        & $\rvdelta=(\delta_1,\dots,\delta_L)$ & Gamma-component delays & Temporal separations between the modes of the positive and negative gamma components. \\
        & $\rvkappa=(\kappa_1,\dots,\kappa_L)$ & Cortical response amplitudes & Peak amplitudes of regional responses obtained by scaling normalized HRF convolutions. \\
        & $\rv{a^-}=(a^-_1,\dots,a^-_L)$ & Negative gamma weights & Relative weights of the negative gamma components in the HRFs. \\
        & $\rv{\xi}_C=(\xi_{C,1},\dots,\xi_{C,L})$ & Cortical HbO weights & Regional HbO fractions; the complementary HbR contributions have opposite sign. \\
        \midrule
        \multirow{8}{*}{$\rvtheta_P$}
        & $\eta$ & Aperiodic PSD amplitude & Scales the aperiodic power-law component of the PSD. \\
        & $R_{\mathrm{phys}}$ & Systemic signal RMS & Target RMS amplitude of the systemic physiology signal before wavelength scaling. \\
        & $\rvalpha=(\alpha_1,\alpha_2,\alpha_3)$ & \makecell[l]{Periodic PSD peak weights\\(Mayer, respiratory, cardiac)} & Relative integrated powers of the three Gaussian spectral peaks. \\
        & $\rvnu=(\nu_1,\nu_2,\nu_3)$ & \makecell[l]{Physiological peak frequencies\\(Mayer, respiratory, cardiac)} & Center frequencies of the three Gaussian spectral peaks. \\
        & $\rho$ & Aperiodic power-law exponent & Controls the frequency-dependent decay of the aperiodic PSD. \\
        & $w_{\mathrm{PSD}}$ & Aperiodic mixture weight & Weights the aperiodic PSD by $w_{\mathrm{PSD}}$ and the periodic PSD by $1-w_{\mathrm{PSD}}$. \\
        & $\xi_P$ & Systemic HbO weight & HbO fraction in the systemic wavelength scaling; the HbR contribution has the same sign. \\
        \midrule
        & $f_\xi(\lambda)$ & Measurement noise fraction & RMS of Gaussian measurement noise relative to the RMS of the normalized signal at wavelength $\lambda$. \\
        \bottomrule
    \end{tabularx}
    \caption{Overview of cortical ($\rvtheta_C$) and systemic physiology ($\rvtheta_P$) parameters governing changes in absorption $\delta\rvmu_a$, together with the separate measurement-noise parameter $f_\xi(\lambda)$. Here, $\rvtheta=(\rvtheta_C,\rvtheta_P)$ and $L$ is the number of active cortical regions.}
    \label{tab:parameters-overview}
\end{table}

\subsection{Theoretical Foundation}\label{sec:observations}

fNIRS exploits the relative transparency of biological tissue in the near-infrared wavelength range $(650-950)\,\si{\nm}$, where measured fluctuations in light absorption can be linked to hemodynamic variations within the probed tissue volume \citep{scholkmann2014review}.
Positioning sources and detectors a few centimeters apart on the scalp enables light to penetrate several centimeters into the tissue, allowing measurements of cortical activity through changes in local hemodynamics associated with neuronal activation. By probing the tissue at multiple wavelengths, oxygenated (HbO) and deoxygenated (HbR) hemoglobin can be distinguished based on differences in their absorption spectra, enabling characterization of their spatiotemporal variations.

In practice, measurements typically follow block-design protocols \citep{luke2021analysis, pinti2019current}, in which a rest period precedes a task or activity period. This design facilitates fNIRS data processing by providing a baseline measurement during rest and enabling analysis of task-related changes in hemoglobin concentrations, such as variations in HbO and HbR during the activity period. Technically, this involves analyzing relative changes between a baseline (unperturbed) detected intensity $\rvphi_0$ (rest) and a perturbed intensity $\rvphi_1(t)$ (task). 

We formulate the simulation of intensity variations using the Rytov approximation. This model, widely used in diffuse optical tomography, relates changes in detected intensity with local variations in the absorption coefficient $\rvdelta\rvmu_a(t;\rvtheta)$ through a sensitivity matrix $\J$ \citep{oleary1996imaging} (see \autoref{sec:sensitivity-matrix}). Let $\mathcal{F}$ denote the forward operator (diffusion solver or Monte Carlo estimator) that maps discretized optical property fields to predicted measurements. After discretizing the domain into $M$ mesh nodes, we represent absorption and scattering as vectors $\rvmu_a,\rvmu_s\in\mathbb{R}^{M}$ and define $\rvphi = \mathcal{F}(\rvmu_a,\rvmu_s)\in\mathbb{R}^{C}$ as the vector of detector measurements.
The Rytov approximation linearizes $\log \rvphi$ with respect to small perturbations in the absorption coefficient $\rvmu_a$ and scattering coefficient $\rvmu_s$ around a baseline $(\rvmu_a^0,\rvmu_s^0)$:
\begin{equation}
    \log \rvphi(\rvmu_a, \rvmu_s) 
    \approx 
    \log \rvphi(\rvmu_a^0, \rvmu_s^0)
    + 
    \frac{\partial \log \rvphi}{\partial \rvmu_a}\bigg|_{\rvmu_a^0} \rvdelta \rvmu_a
    +
    \frac{\partial \log \rvphi}{\partial \rvmu_s}\bigg|_{\rvmu_s^0} \rvdelta \rvmu_s.
\end{equation}
\review{Notably, we use the Rytov approximation for a balance of accuracy and computational efficiency as compared to first-order or higher-order Born approximations \citep{oleary1996imaging,sassaroli2024revisiting}.}
As customary in the field, scattering is kept constant (i.e., $\rvdelta\rvmu_s \approx 0$). Explicating the dependence of $\rvdelta \rvmu_a$ on the tunable parameters $\rvtheta$ yields the standard Rytov formulation:
\begin{equation}\label{eq:rytov}
    \log \Delta \rvphi(\lambda, t; \rvtheta) = \log\!\left(\frac{\rvphi_1(\lambda, t; \rvtheta)}{\rvphi_0(\lambda)}\right) 
    \approx 
    \J(\lambda) \,\rvdelta\rvmu_a(\lambda, t; \rvtheta),
\end{equation}
where $\rvphi_0 \equiv \rvphi(\rvmu_a^0, \rvmu_s^0)$ is the unperturbed intensity, $\rvphi_1(t; \rvtheta) \equiv \rvphi(\rvmu_a, \rvmu_s)$ the perturbed intensity, and $\J$ is the Jacobian of $\log \rvphi$ with respect to $\rvmu_a$ evaluated at $\rvmu_a^0$. Rearranging gives a direct expression for the predicted intensity:
\begin{equation}
    \rvphi_1(\lambda, t; \rvtheta) \approx \rvphi_0 \exp\left( \J(\lambda) \,\rvdelta\rvmu_a(\lambda, t; \rvtheta) \right),
\end{equation}
which forms the basis of our simulation model. 
From this simulated intensity, we compute the change in optical density ($\Delta\mathrm{OD}$) by log-normalizing with respect to the average baseline intensity over a pre-stimulus period $\bar{\rvphi}_{\mathrm{base}}$:
\begin{equation}
    \Delta\mathrm{OD}(\lambda, t; \rvtheta) =
    \log \bar{\rvphi}_{\mathrm{base}}(\lambda) -
    \log \rvphi_1(\lambda, t; \rvtheta),
\end{equation}
where 
\begin{equation}
    \bar{\rvphi}_{\mathrm{base}}(\lambda) = \frac{1}{T_{\mathrm{base}}}\int_{0}^{T_{\mathrm{base}}}\!\rvphi_1(\lambda, t; \rvtheta)\,\mathrm{d}t
\end{equation}
is the average intensity during the baseline window of length $T_{\mathrm{base}}$.
Of note, $\bar{\rvphi}_{\mathrm{base}}$ is computed as the mean pre-stimulus intensity for each channel and is different from the simulated unperturbed measurement vector $\rvphi_0$, which is generated via MMC.
\review{
To convert from $\Delta \mathrm{OD}$ to changes in HbO and HbR concentration, we use the Modified Beer-Lambert Law \citep[MBLL;][]{baker2014modified}:
\begin{equation}
    \begin{pmatrix}
    \Delta\mathrm{HbO}(t; \rvtheta)\\
    \Delta\mathrm{HbR}(t; \rvtheta)
    \end{pmatrix}
    =
    \frac{1}{d_{\mathrm{RD}}}
    \begin{pmatrix}
    \mathrm{DPF}(\lambda_1)\varepsilon_{\mathrm{HbO}}(\lambda_1) &
    \mathrm{DPF}(\lambda_1)\varepsilon_{\mathrm{HbR}}(\lambda_1)\\
    \mathrm{DPF}(\lambda_2)\varepsilon_{\mathrm{HbO}}(\lambda_2) &
    \mathrm{DPF}(\lambda_2)\varepsilon_{\mathrm{HbR}}(\lambda_2)
    \end{pmatrix}^{-1}
    \begin{pmatrix}
    \Delta \mathrm{OD}(\lambda_1,t; \rvtheta)\\
    \Delta \mathrm{OD}(\lambda_2,t; \rvtheta)
    \end{pmatrix}.
\end{equation}
where $\epsilon(\lambda)$ are wavelength-dependent molar extinction coefficients of HbO and HbR \citep{wray1988characterization, prahl1999hemoglobin}, $d_{\mathrm{RD}}$ is the distance from source to detector in $\mathrm{mm}$, and $\mathrm{DPF}(\lambda)$ is a differential pathlength factor \citep{chiarelli2019differential}. Noteworthy, $\varepsilon(\lambda)$ were converted from decadic to natural-logarithmic absorption coefficients and from $\mathrm{M}^{-1}\mathrm{cm}^{-1}$ to $\mathrm{M}^{-1}\mathrm{mm}^{-1}$ by multiplication with $\ln(10)/10$ to maintain consistency with the natural-logarithmic intensity model.
}

\subsection{Generating the Sensitivity Matrix}\label{sec:sensitivity-matrix}

To model relative changes between the baseline detected intensity $\rvphi_0$ (rest) and a perturbed intensity $\rvphi_1(t)$ (task) in measured intensity, light propagation models can be employed \citep{dehghani2009near, fang2010mesh}. 
In this context, Monte Carlo methods are widely regarded as the most accurate approach for modeling photon transport in complex biological tissues across a broad range of optical properties and measurement configurations. However, these methods are often computationally expensive, particularly when simulating deep tissue propagation, which requires large numbers of photon packets, or when repeated simulations are needed \citep{chen2011comparison}. 

To address this challenge, we leverage the Mesh-based Monte Carlo toolbox \citep[MMC;][]{fang2010mesh}, which enables efficient computation of three-dimensional tissue sensitivity profiles $\J$ of measured intensity $\rvphi_1$ to changes in the absorption coefficient $\rvmu_a$ \citep{yao2018direct,nizam2022monte}.
MMC simulates light transport through a tetrahedral mesh of biological tissue, providing a flexible framework for generating subject- or atlas-based forward models. Its practical applicability has improved significantly in recent years due to advances in computational power and GPU parallelization \citep{fang2019graphics}.

The inputs to the MMC simulations consist of three main components: anatomical information, experimental configuration, and tissue-specific optical properties. 
First, for the anatomical model, we followed the procedure described by \cite{zimeo2018fnirs} to generate a high-resolution segmented tetrahedral brain mesh based on the Colin27 atlas \citep{holmes1998colin27}.

Second, the experimental configuration is defined by the optode montage (\autoref{fig:Montage}). The montage comprises both long- and short-separation channels, where each channel corresponds to a source-detector pair. Short-separation channels are primarily sensitive to superficial/systemic signals, whereas long-separation channels are more sensitive to deeper, cortical activations, although typically at the cost of a lower signal-to-noise ratio (SNR). Different montages are used for the different experimental protocols described in \autoref{sec:experimental-data} \citep{yucel2015specificity, kamat2023assessment}. 

\begin{table}[!ht]
    \centering
    \begin{tabular}{ll}
        \toprule
        \textbf{Dimension} & \textbf{Description}\\
        \midrule
        $C$ & Number of channels\\
        $R$ & Number of sources\\
        $D$ & Number of detectors\\
        $L$ & Number of active cortical regions\\
        $M$ & Number of mesh nodes\\
        $E$ & Number of mesh elements\\
        \bottomrule
    \end{tabular}
    \caption{Overview of number of dimensions of key quantities in the forward model.}
    \label{tab:dim-abbrev}
\end{table}

Before running the simulations, the digitized probe geometry is registered to the tetrahedral head mesh. The 3D optode coordinates are first rotated into RAS (Right, Anterior, Superior) orientation and then projected onto the mesh surface using the iterative closest point (ICP) algorithm \citep{besl1992method}. The source directions are assumed to point from the optode locations toward the centroid of the mesh. The optodes are embedded one mean free path beneath the mesh surface \citep{fang2010mesh}.

Third, the tissue-specific optical properties used in the simulations include the absorption coefficient $\mu_a$, the scattering coefficient $\mu_s$, the anisotropy factor $g$, and the refractive index $m$. Their values were obtained from existing literature and are summarized in \autoref{tab:optical-properties}.

Simulating photon propagation through the mesh yields the detected baseline fluence $\rvphi_0$, as well as the spatial distribution of fluence inside the brain mesh in the form of a Green's function, assuming a unit-intensity light source \citep{fang2009mcx}. We then compute the sensitivity matrix for measurement channel source-detector pairings $\J \in \sR^{C \times M}$, using the adjoint method \citep{yao2018direct}.

More specifically, we first run $R$ simulations, one for each source, to obtain the Green's function $\mathbf{G}^R \in \sR^{R \times M}$, which describes the normalized fluence of each mesh node for each source. These simulations also provide the detected intensity for all montage-defined source-detector pairs, which we normalize via the area of the detector $A_d$, and the total number of simulated photons $N_p$, denoted by $\mathbf{G}^{RD} \in \sR^{C}$. We then use the reciprocity of light transport and run $D$ additional simulations, placing a virtual source at each detector. This yields $\mathbf{G}^D \in \sR^{D \times M}$, which describes the normalized fluence of each mesh node for each detector.
\review{
The value of the Jacobian $\J_{i,m}$ at mesh node $m = 1 , \dots, M$ and for each measurement channel $i=1, \dots , C$, defined by the source-detector pair $(r_i, d_i)$, is calculated as
\begin{equation}
    \J_{i,m} = - V_m\frac{\mathbf{G}^R_{r_i,m} \odot \mathbf{G}^D_{d_i,m}}{\mathbf{G}^{RD}_{i}}.
\end{equation}
Here, $\mathbf{G}^R_{r_i,m}$ denotes the fluence at node $m$ due to source $r_i$, and $\mathbf{G}^D_{d_i,m}$ denotes the corresponding adjoint fluence for detector $d_i$. The operator $\odot$ denotes element-wise multiplication. The nodal volume $V_m \in \sR$ is defined as
\begin{equation}
    V_m = \frac{1}{4}\sum_{\tau \in \mathcal{T}(m)} V_\tau,
\end{equation}
where $\mathcal{T}(m)$ is the set of tetrahedral elements sharing node $m$, and $V_\tau$ is the volume of tetrahedral element $\tau$, calculated from its four vertex coordinates.} 

\subsection{Modeling Change in Absorption}\label{sec:deltamua}

The total change in absorption coefficient for wavelength $\lambda$ across all $M$ mesh nodes is modeled as a weighted sum of cortical and systemic contributions that are parameterized via $\rvtheta$ (see \autoref{tab:parameters-overview}):
\review{\begin{equation} \label{eq:deltamua}
    \rvdelta\rvmu_a(\lambda, t;\rvtheta) = 
    \rvdelta\rvmu_C(\lambda, t; \rvtheta_C) + 
    \rvdelta\rvmu_{P}(\lambda ,t; \rvtheta_P),
\end{equation}}
where $\rvdelta\rvmu_C(\lambda, t;\rvtheta_C)$ is the absorption change due to \emph{cortical hemodynamic activity}, and $\rvdelta\rvmu_{P}(\lambda, t;\rvtheta_P)$ represents absorption changes caused by \emph{systemic physiology} (e.g., heart rate, respiration, blood pressure).

\subsubsection{Cortical Hemodynamic Activity}\label{sec:cortical-activations}
\begin{figure}[t]
    \centering
    \includegraphics[width=1.0\textwidth]{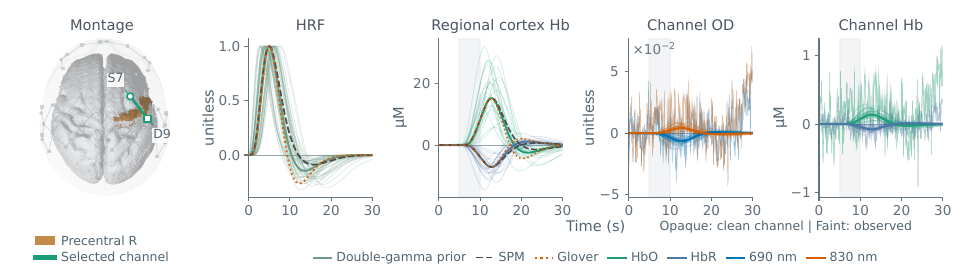}
    \caption{\review{Example simulations of the HRF family with resulting regional cortical concentration changes and implied synthetic measurements (change in optical density or in hemoglobin concentration) after adding physiological contributions and projection into channel space. Notably, we parameterize the HRF family to overlap with standard SPM or Glover models. Furthermore, the simulator's modular design allows projecting cortical concentration changes in isolation, creating pairs of clean cortical and physio-contaminated channel recordings.}}
    \label{fig:cortical_contribution}
\end{figure}

The cortical contribution is modeled as a linear combination of \review{\textit{regional cortical concentration changes} $\delta \rvmu_C^{HbO}(\lambda, t; \rvtheta_C)$ and $\delta \rvmu_C^{HbR}(\lambda, t; \rvtheta_C)$} (see \autoref{fig:cortical_contribution}; third column):
\begin{align}\label{eq:absorption_change}
    \rvdelta \rvmu_C(\lambda, t; \rvtheta_C) &= \delta \rvmu_C^{HbO}(\lambda, t; \rvtheta_C) + \delta \rvmu_C^{HbR}(\lambda, t; \rvtheta_C)\\
    &=\mathbf{A} \bigl( \rv{s}_C(\lambda;\rvtheta_C) \odot \rvc(t; \rvtheta_C) \bigr),
\end{align}
where $\mathbf{A} \in \sR^{M \times L}$ is a map of activated regions, $\rvc(t; \rvtheta_C)\in \sR^{L}$ describes their temporal amplitudes, and $\rv{s}_C(\lambda; \rvtheta_C)\in \sR^{L}$ is a vector of wavelength-dependent scaling factors.
\review{
Each column $\ell = 1, \dots L$ of $\mathbf{A}$ is a binary spatial map corresponding to a functional brain region defined by the Automated Anatomical Labeling Atlas 3 (AAL3) \citep{rolls2020automated} which was coregistered using SPM12 \citep{friston2003statistical}. An activated region is set to one at the mesh nodes coinciding with their AAL3 labeled locations, while inactive regions are set to zero. This design choice allows activating specific functional regions of the brain as motivated by the literature (\autoref{sec:experimental-protocol}), while remaining easily extendable to alternative parcellations.
}

\review{
The scaling factors are computed as
\begin{align}
    \rv{s}_C(\lambda;\rvtheta_C) &= \rv{s}_C^{HbO}(\lambda;\rvtheta_C) + \rv{s}_C^{HbR}(\lambda;\rvtheta_C)\\
    &= \rv{\xi}_C \, \epsilon_{\mathrm{HbO}}(\lambda)-(\mathbf{1} - \rv{\xi}_C) \, \epsilon_{\mathrm{HbR}}(\lambda),
\end{align}
where $\rv{\xi}_C$ is part of $\rvtheta_{C}$ and controls the relative weight of $\Delta$HbO with respect to $\Delta$HbR and $\epsilon_{\cdot}(\lambda)$ are molar extinction coefficients of HbO and HbR \citep{wray1988characterization, prahl1999hemoglobin}.
}

\review{
\paragraph{Latent cortical concentration changes}
For simulating regional concentration changes $\rvc(t; \rvtheta_C)$ with varying amplitudes, shapes, and time-to-peak dependent on the experimental stimulus, we modularize generation for each activation region $\ell$ into three main components: a positive peak variable $\kappa_\ell$ in molar units (M), a hemodynamic response function (HRF) family, and the stimulus vector $x(t)$.
In this way, we first compute a latent cortical concentration $q_\ell$ by convolving the experimental design vector $x(t)$ with a unit-positive-peak HRF $h_\ell(t)$ followed by scaling the normalized response $\tilde{q}_\ell(t)$ with $\kappa_\ell$:
\begin{equation}\label{eq:conv}
    q_\ell(t) = (x * h_\ell)(t), \qquad
    \tilde{q}_\ell(t) = \frac{q_\ell(t)}{\max_{t'} q_\ell(t')}, \qquad
    c_\ell(t; \rvtheta_{C,\ell}) = \kappa_\ell\, \tilde{q}_\ell(t),
    \qquad \ell = 1,\ldots,L.
\end{equation}
}

\paragraph{Hemodynamic Response Function (HRF)}
\review{ 
The HRF $h_\ell(t)$ models the stereotypical shape of the cortical response (see \autoref{fig:cortical_contribution}; second column). For short stimuli, double-gamma functions are typically used \citep{lindquist2009modeling}.
We define the gamma density in terms of a dimensionless shape $a>0$ and a scale $d > 0$:
\begin{align}
    g(t;a,d) &= \frac{t^{a-1}e^{-t/d}}{\Gamma(a)d^a} \\
    \tilde h_\ell(t) &=
    g\!\left(t;1+\frac{\tau_\ell}{d^+_\ell},d^+_\ell\right)
    -a^-_\ell\,
    g\!\left(t;1+\frac{\tau_\ell+\delta_\ell}{d^-_\ell},d^-_\ell\right) \label{eq:double_gamma_shape}.
\end{align}
Here $\tau_\ell>0$ is the time of the positive gamma component's mode, $d^+_\ell>0$ and $d^-_\ell>0$ are the positive- and negative-component gamma scales (dispersions), $\delta_\ell>0$ is the separation between their modes, and $a^-_\ell\ge0$ is the dimensionless weight of the negative gamma component. The timing and dispersion parameters are in seconds. The component modes are $\tau_\ell$ and $\tau_\ell+\delta_\ell$; these need not coincide exactly with the extrema of their difference. An HRF is modeled for each selected anatomical region $\ell$ of the AAL3-segmented mesh. 
In summary the shape of the HRF is governed by $(\tau_\ell,d^+_\ell,d^-_\ell,\delta_\ell,a^-_\ell) \subset \rvtheta_{C,\ell}$, and to ensure $\kappa_\ell$ solely controls the regional response peak, the resulting, unitless HRF shape is normalized:
\begin{equation}\label{eq:hrf_unit_peak}
    h_\ell(t)=\frac{\tilde h_\ell(t)}{\max_{t'} \tilde h_\ell(t')}.
\end{equation}}

As shown in \cite{lindquist2009modeling} the addition of temporal and dispersion derivatives may improve HRF modeling. Our simulator allows for the optional inclusion of these derivatives in modeling the temporal response. Aside from the variable double gamma HRF model with sampled time to peak, and HRF width values, we also implement the Glover HRF model \citep{glover1999deconvolution} and the SPM HRF model \citep{lindquist2009modeling} (see \autoref{fig:cortical_contribution}). For all analyses in this paper, the variable double gamma HRF model is used.

\paragraph{Design Vector}
A binary time-dependent stimulus state $x(t)\in \{0,1\}$ indicates when a subject is presented with an experimental event (e.g., 1 = ON, 0 = OFF) based on the experimental task.
For example, in a block-design finger-tapping task with 5 seconds of tapping and 20 seconds of rest \citep{yucel2015specificity}, the design vector for a given trial would look like:
\begin{equation}
    \rvx = [\underbrace{1,\dots,1}_{5\text{ s}},\underbrace{0,\dots,0}_{20\text{ s}}].
\end{equation}
Thus, convolving the HRF with the design vector (\autoref{eq:conv}) ensures that the simulated activation follows the timing of the experimental protocol (\autoref{sec:experimental-data}).

\subsubsection{Systemic Physiology}\label{sec:systtemic-physiology}

\begin{table*}[t]
    \centering
    \resizebox{\textwidth}{!}{%
    \begin{tabular}{p{7cm}cp{7cm}}
        \toprule
        \textbf{Physiological Origin} & \textbf{Frequency Band} $[\si{\hertz}]$ & \textbf{Reference} \\
        \midrule
        Cardiac Activity & $.60-2.5$ & \cite{leamy2011functional,hocke2018automated,nguyen2018adaptive} \\
        Respiratory Activity & $.20-.60$ & \cite{hocke2018automated,nguyen2018adaptive} \\
        Blood Pressure Fluctuations (Mayer Waves) & $.06-.14$ & \cite{nguyen2018adaptive,luke2021characterization} \\
        \bottomrule
    \end{tabular}%
    }
    \caption{Systemic physiology frequency bands used to construct informative priors over the systemic physiology parameters of our simulator.}
    \label{tab:super-phys-freq-band}
\end{table*}

We model changes in absorption due to systemic physiology $\rvdelta \rvmu_{P}$ (see \autoref{fig:parameter_variations}) in the frequency domain, construct a complex-valued frequency spectrum, and subsequently transform it into the time domain via the inverse Fourier transform \citep{mcsharry2003dynamical}:
\review{
\begin{align}
    \rvdelta\tilde\rvmu_P(t;\rvtheta_P) &= \gF^{-1}\left[\rvdelta M_P(f;\rvtheta_P)\right], \\
    \rvdelta\rvmu_P(\lambda,t;\rvtheta_P) &= \rv{s}_P(\lambda;\rvtheta_P) \odot \rvdelta\tilde\rvmu_P(t;\rvtheta_P).
\end{align}
}
\review{The wavelength-dependent scaling $\rv{s}_P(\lambda;\rvtheta_P)$ is defined as: 
\begin{equation}
    \rv{s}_P(\lambda;\rvtheta_P) = \rv{\xi}_P \, \epsilon_{\mathrm{HbO}}(\lambda)+(\mathbf{1} - \rv{\xi}_P) \, \epsilon_{\mathrm{HbR}}(\lambda),
\end{equation}
where $\rv{\xi}_P$ defines the relative contributions of HbO and HbR.}
The complex frequency-domain representation $\rvdelta M_P$ is defined as
\begin{equation}
    \rvdelta M_P(f; \rvtheta_P) =
    \sqrt{\mathrm{PSD}(f;\rvtheta_P)\,\frac{df}{2}}\;
    e^{j \beta(f)},
\end{equation}
where $df$ is the frequency resolution (inversely proportional to the time-series length), 
$j$ the imaginary unit, and $\beta(f)\sim \mathcal{U}[0,2\pi)$ is a randomly sampled phase ensuring proper phase scrambling.
\review{The power spectral density (PSD) is modeled as the weighted sum of an aperiodic power-law component and periodic Gaussian peaks corresponding to physiological oscillations:
\begin{align}
    \widetilde{\mathrm{PSD}}(f;\rvtheta_P) &= w_{\mathrm{PSD}}\mathrm{PSD}_{\mathrm{aperiodic}}(f;\rvtheta_P) + (1-w_{\mathrm{PSD}})\mathrm{PSD}_{\mathrm{periodic}}(f;\rvtheta_P), \\
    \mathrm{PSD}(f;\rvtheta_P) &= R_{\mathrm{phys}}^2\frac{\widetilde{\mathrm{PSD}}(f;\rvtheta_P)}{\int_0^{r/2}\widetilde{\mathrm{PSD}}(f';\rvtheta_P)\,\mathrm{d}f'}.
\end{align}
Here, $w_{\mathrm{PSD}}\in[0,1]$ weights the aperiodic component, $r$ is the sampling rate, and $R_{\mathrm{phys}}$ is the target RMS of the systemic $\Delta\mathrm{HbT}$ (total hemoglobin) waveform before wavelength scaling.}
The \textit{aperiodic component} accounts for broadband background fluctuations and is modeled as a 
power-law ($1/f$) process with a low-frequency stabilizer:
\begin{equation}
    \mathrm{PSD}_{\mathrm{aperiodic}}(f;\rvtheta_P) =
    \frac{\eta}{(f + \tilde{f})^{\rho}} + u,
\end{equation}
where $\eta$ is the spectral power offset, $\rho$ is the $1/f$ exponent controlling the slope, 
and $\tilde{f} > 0$ is a small stability constant that prevents divergence as $f \to 0$. 
The term $u \sim \mathcal{N}(0,\sigma_u^2)$ represents additive Gaussian noise with a small scale ($\sigma_u \approx 0.1$), which creates variability across frequency bins.

Differently, the \textit{periodic component} models rhythmic oscillations in three characteristic frequency bands 
(e.g., cardiac, respiratory, and Mayer waves; see \autoref{tab:super-phys-freq-band}) as a 
mixture of Gaussian peaks:
\begin{equation}
    \mathrm{PSD}_{\mathrm{periodic}}(f;\rvtheta_P) =
    \sum_{b=1}^{3}
    \frac{\alpha_b}{\sqrt{2\pi}\,\sigma_b}
    \exp \left[
        -\frac{1}{2}
        \left(\frac{f-\nu_b}{\sigma_b}\right)^2
    \right],
\end{equation}
where $\alpha_b$ is the amplitude, $\nu_b$ the central frequency, and $\sigma_b \in \{0.015, 0.02, 0.08\}$ the spectral width of each component.  
The values of $\sigma_b$ are heuristically chosen and can be tuned to match subject-specific physiology.
\review{Further frequency components \citep[e.g., very low frequency oscillations, VLFO;][]{bicciato2021increase} can easily be incorporated.}

\subsection{Measurement effects}\label{sec:measurement-effects}
\subsubsection{External Artifact Generation}\label{sec:artifacts}
\begin{figure}[t]
    \centering
    \includegraphics[width=1.0\textwidth]{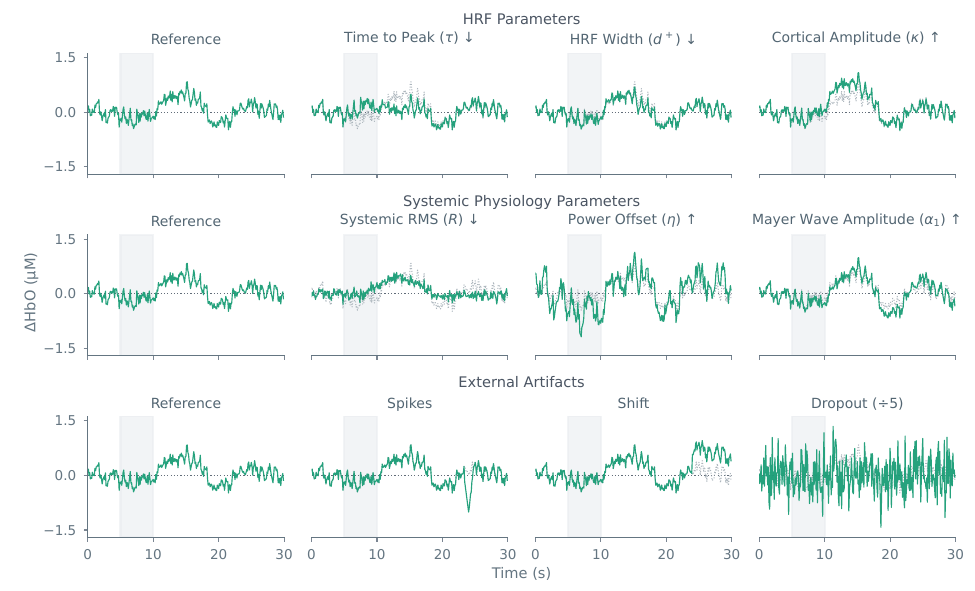}
    \caption{The impact of HRF parameters (\textit{top}) systemic physiology parameters (\textit{middle}) and external artifacts (\textit{bottom}) on the change of HbO over time. The shaded region indicates the stimulus period. Dropout values (bottom right) are downscaled by a factor of five for a more dynamic visualization range.}
    \label{fig:parameter_variations}
\end{figure}

To generate synthetic data representative of artifacts commonly observed in fNIRS, we explicitly model motion artifacts (i.e., spikes and baseline shifts) and apply them to the simulated raw intensity time series $\rvphi_1(t;\rvtheta)\in\mathbb{R}^{C}$ \citep{gao2022deep,barker2013autoregressive} (see \autoref{fig:parameter_variations}). 
Artifacts are generated independently for each channel and then combined.

Let $\rvpsi_s(t)\in\mathbb{R}^{C}$ denote the spike artifact vector and $\rvpsi_b(t)\in\mathbb{R}^{C}$ the baseline-shift artifact vector at time index $t$. The corrupted intensity is defined as \begin{equation}\label{eq:artifact_application} \tilde{\rvphi}_1(t;\rvtheta) = \rvphi_1(t;\rvtheta) +
\mathrm{diag} \bigl(\sigma(\rvphi_1(t;\rvtheta))\bigr)\,\Bigl(\rvpsi_s(t)+\rvpsi_b(t)\Bigr),
\end{equation}
where $\sigma(\rvphi_1(\cdot;\rvtheta))\in\mathbb{R}^{C}$ is the vector of channel-wise standard deviations of the raw intensity time series (computed per channel over $t=1,\dots,T$), and $\mathrm{diag}(\cdot)$ converts it to a diagonal scaling matrix. This scaling expresses artifact magnitudes relative to each channel’s native intensity variability.
To prevent unrealistically low (or negative) intensity values after corruption, we clip elementwise at a small heuristically chosen floor $u>0$:
\begin{equation}\label{eq:intensity_floor}
\tilde{\rvphi}_1(t;\rvtheta) \leftarrow \operatorname{\max} \bigl(\tilde{\rvphi}_1(t;\rvtheta),\,u\bigr).
\end{equation}

\paragraph{Spike Generation}

To simulate temporary optode displacements, we model spike artifacts as exponentially decaying pulses centered at randomly selected time points, independently for each channel. Let $p_s$ denote the per-sample probability of a spike center, obtained from a per-second probability $P_s$ via
\begin{equation}
    p_s = \frac{P_s}{r},
\end{equation}
where $r$ is again the sampling rate (Hz). For each channel $c\in\{1,\dots,C\}$ and time index $t\in\{1,\dots,T\}$, we define the random set of spike-center times in channel $c$ as $\mathcal{T}_{s,c} = \{t : z^{(s)}_{t,c}=1\}$, with $z^{(s)}_{t,c} \sim \mathrm{Bernoulli}(p_s)$.
For each $\tau \in \mathcal{T}_{s,c}$ we sample a half-width $w_{\tau,c}$ and signed amplitude $a_{\tau,c}$ as
\begin{equation}
    w_{\tau,c} \sim \mathrm{DiscreteUniform}\bigl(\lfloor r/4\rfloor,\;\lfloor 2r\rfloor\bigr),
    \qquad
    a_{\tau,c} \sim \mathrm{Uniform}\bigl(-a_s/2,\;a_s/2\bigr),
\end{equation}
with a heuristically determined scaling parameter $a_s>0$. The sum of spike artifacts for channel $c$ is described by the following:
\begin{equation}\label{eq:spike_superposition}
    \psi_{s,c}(t)
    =
    \sum_{\tau \in \mathcal{T}_{s,c}}
    a_{\tau,c}\,
    \exp\left(-\frac{|t-\tau|}{w_{\tau,c}}\right)\,
    \mathbb{I}\{|t-\tau|\le w_{\tau,c}\}.
\end{equation}
Stacking across channels gives the spike vector $\rvpsi_s(t)=(\psi_{s,1}(t),\dots,\psi_{s,C}(t))^\top$.

\paragraph{Baseline Shift Generation}

To simulate permanent changes in optode coupling, we model baseline shifts as step-like offsets with randomly selected onset times, independently for each channel.
Let $p_b$ denote the per-sample probability of a baseline-shift onset, obtained from a per-second probability $P_b$ via
\begin{equation}
    p_b = \frac{P_b}{r},
\end{equation}
where $r$ is the sampling rate (Hz). For each channel $c\in\{1,\dots,C\}$ and time index $t\in\{1,\dots,T\}$, we define the (random) set of onset times in channel $c$ as $\mathcal{T}_{b,c} = \{t : z^{(b)}_{t,c}=1\}$, with $z^{(b)}_{t,c} \sim \mathrm{Bernoulli}(p_b)$.
For each $\tau \in \mathcal{T}_{b,c}$, we sample a signed step amplitude $a_{\tau,c} \sim \mathrm{Uniform}\bigl(-a_b/2,\;a_b/2\bigr)$
with scaling parameter $a_b>0$. Accordingly, the sum of baseline-shift artifacts for channel $c$ is given by:
\begin{equation}\label{eq:baseline_shift_superposition}
    \psi_{b,c}(t) =
    \sum_{\tau \in \mathcal{T}_{b,c}}
    a_{\tau,c}\,
    \mathbb{I}\{t\ge \tau\},
\end{equation}
where stacking across channels gives the baseline-shift vector $\rvpsi_b(t)=(\psi_{b,1}(t),\dots,\psi_{b,C}(t))^\top$.

\review{\paragraph{Design-Dependent Artifact Generation}
In experimental paradigms involving speech or other task-related movements, motion artifacts may systematically covary with the experimental design, potentially introducing spurious task-evoked responses \citep{brigadoi2014motion,novi2020speech}. To accommodate such dependencies, we allow any artifact parameter (e.g., occurrence probabilities or amplitudes) to depend on the experimental design vector $x(t)$ introduced in \autoref{sec:cortical-activations}.
In general, this can be achieved by regressing the respective parameter on $x(t)$, subject to its domain constraints.
For example, let $k \in \{s,b\}$ index spike and baseline-shift artifacts. Their event probabilities can be modeled as
\begin{equation}\label{eq:task_artifacts}
    p_k(t) = \frac{P_k^{(0)} + \beta_k x(t)}{r}, \qquad z_{t,c}^{(k)} \sim \mathrm{Bernoulli}(p_k(t)),
\end{equation}
where $P_k^{(0)}$ denotes the baseline per-second event probability and $\beta_k$ controls its task-dependent change.
Similarly, parameters governing the amplitude distributions can depend on $x(t)$, allowing both artifact occurrence and magnitude to covary with experimental conditions.
Setting the regression coefficients to zero recovers the stimulus-independent model described above.}

\paragraph{Channel Dropout}
Finally, to model channel dropout, we assume a random probability $P_d$ (determined heuristically for each dataset) that both waveforms of any given channel are replaced by Gaussian noise with mean equal to that channel's mean and amplitude $A_d$. Due to the possibility of random noise producing negative intensity values, any noisy channel with a minimum below a heuristically determined threshold, $u_d$, is rescaled so that the channel mean remains unchanged while the minimum equals the threshold.

\subsubsection{Measurement noise}\label{sec:measurement-noise}
\review{
To represent measurement noise, Gaussian random noise $\xi$ is added to the Rytov-normalized measurement vector $\log \Delta \rvphi$ for each channel $c$. Unit Gaussian noise $\tilde{\xi}_{c, \lambda} \sim \gN(0, 1)$ is rescaled to the root mean square (RMS) of the measurement channel:
\begin{equation}
    \xi_{c, \lambda} = f_\xi(\lambda)\operatorname{RMS}\left(\log \Delta\rvphi\left(c, \lambda\right)\right) \frac{\tilde{\xi}_{c, \lambda}}{\operatorname{RMS}(\tilde{\xi}_{c, \lambda})},
\end{equation}
where $f_\xi(\lambda)$ is a wavelength-dependent noise fraction parameter.}

\section{Experiments} \label{sec: Methods}
\begin{figure}[t]
    \centering
    \includegraphics[width=1.0\textwidth]{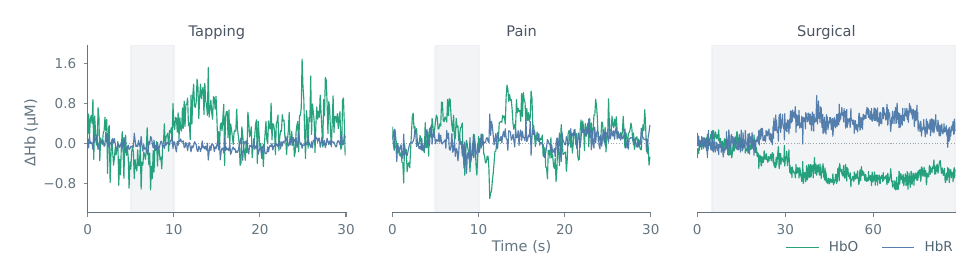}
    \caption{Examples of simulated chromophore changes projected into channel space across different tasks (from \textit{left to right}): finger tapping, pain assessment, and surgical skill assessment, using the specifically calibrated priors described in the appendix (\autoref{FingerTapping_Params} and \autoref{Surgical_Params}). The stimulus period is indicated by the gray shaded region.}
    \label{fig:parameter_variations_tasks}
\end{figure}
\subsection{Experimental Data}\label{sec:experimental-data}
The experimental data used to validate our simulator come from the publicly available finger-tapping and electrical pain assessment datasets \citep{yucel2015short} and the sham condition of the long-term tES surgical pattern cutting dataset \citep{kamat2023assessment}. For the surgical skill assessment dataset, we use trials between 30 and 98 seconds, since anything below 30 seconds is likely to be a mistake (e.g., a trigger pressed twice), and anything above 98 seconds is considered an outlier (based on the protocol design). Here, we use only the sham (control) condition to assess our proposed method's ability to model differences in experimental hardware, subjects, and task duration. \review{To support efficient generation of large synthetic training datasets for downstream machine learning, we demonstrate dataset-specific prior calibration rather than relying on a single broad prior that would generate many recordings outside the experimental setting of interest.}
For ease of analysis, we crop all experimental and simulated trials to 30 seconds. Below, we provide brief descriptions of the experimental hardware, subjects, and protocols.

\subsubsection{Hardware and Equipment}
\begin{figure*}[t]
    \centering
    \includegraphics[width=1.0\textwidth]{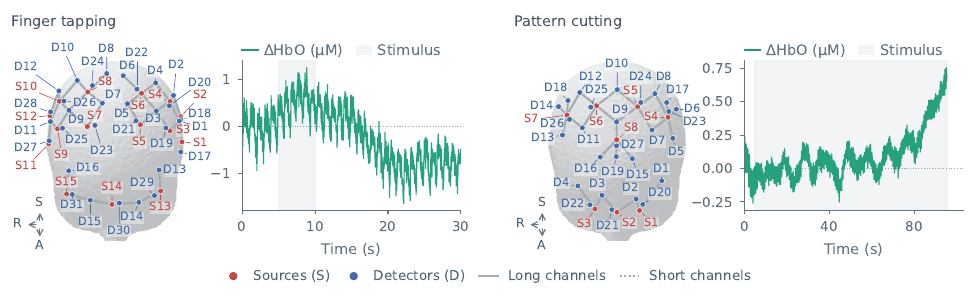}
    \caption{
    Registered optodes of the experimental montages for finger tapping and pain assessment \citep[\textit{left};][]{yucel2015short}, and surgical skill assessment \citep[\textit{right};][]{kamat2023assessment} datasets. Examples of measured change in HbO are shown for two channels with overlaid stimulus period windows.}
    \label{fig:Montage}
\end{figure*}

The finger tapping dataset used a continuous-wave 32-channel near-infrared spectrometer that delivered light at 690 nm and 830 nm (CW6 system, TechEn Inc., MA, USA).
The experimental montage consisted of 15 sources, 16 long-separation detectors, and 15 short-separation detectors, with source-detector separations of approximately 3 cm and 0.8 cm for long- and short-separation channels, respectively. The probe covers the frontal cortex and portions of the somatosensory and motor areas across both hemispheres (\autoref{fig:Montage}, \textit{top left}). 

The surgical skill assessment dataset used a continuous-wave near-infrared spectrometer that delivered light at 760 nm and 850 nm (NIRScout system, NIRx, Berlin, Germany). The experimental montage comprised 8 sources, 19 long-separation detectors, and 8 short-separation detectors, with source-detector separations of approximately 3 cm and 0.8 cm for the 28 long- and 8 short-separation channels, respectively. The probe covers the prefrontal cortex, primary motor cortex, and supplementary motor area across both hemispheres (\autoref{fig:Montage}, \textit{right}). Transcranial electrical stimulation (tES) was delivered via a StarStim device (StarStim, Neuroelectronics, Spain).

\subsubsection{Subjects and Experimental Protocol}\label{sec:experimental-protocol}
The finger tapping/electrical stimulus experimental dataset consists of 12 healthy, right-handed males ($25 \pm 5$ years old) with no history of neurological trauma or psychiatric disorders. Subjects completed a left-hand finger-tapping task with a block design of 5 s of stimulus followed by 25 s of rest, repeated 12 times for a total of 6 minutes, during which fNIRS recordings were taken, resulting in an included 144 trials. Example simulated trials for this dataset, along with a detailed list of model parameter distributions, are provided in  \autoref{FingerTapping_Params}, with a simulated example trial shown in \autoref{fig:parameter_variations_tasks}.

The surgical skill assessment sample consists of 7 healthy, right-handed subjects with a mean age of 21 $\pm$ 1. The surgical skill assessment task performed was the FLS pattern-cutting task, in which subjects attempted to cut a pre-marked circle from a gauze sheet within an FLS box simulator \citep{kamat2023assessment}. Before beginning the training, subjects watched a tutorial video followed by a practice trial. In the following 10 training days, subjects underwent sham tES, in which a transcranial electrical stimulation ramped up from 0 to 1 mA over 30 s, then ramped back down to 0 mA at both the beginning and end of a 10-minute stimulation period. Afterward, subjects performed repeated trials of the pattern-cutting task for 30 minutes, with 30-second rest intervals between trials. This resulted in 3-10 repetitions of pattern cutting per training period, with the training periods occurring on 10 consecutive days after the introductory day. On the day following the final training day, subjects performed a post-test consisting of 3 trials of the pattern-cutting FLS task. 30 seconds after the final post-test trial, subjects performed a transfer learning task, where they removed a marked circle from a layer of \textit{ex vivo} porcine abdominal tissue. Fourteen days after this post-test, subjects performed a retention test identical to the post-test, consisting of 3 pattern-cutting trial repetitions followed by 1 transfer-learning trial. After removing trials outside of the specified range, a total of 735 trials were considered. Example simulated trials for this dataset, along with a detailed list of model parameter distributions, are provided in \autoref{Surgical_Params}, with a simulated example trial shown in \autoref{fig:parameter_variations_tasks}.

\subsection{Model Comparison}
\label{sec:benchmarks}

To evaluate the proposed Jacobian-based simulator, we compare it against two alternative simulation approaches. The first is an autoregressive (AR) simulator implemented in existing fNIRS analysis toolboxes. The second is an ablated Jacobian-free version of our simulator that retains the same generative components but omits the Jacobian-based spatial projection.
\review{Here, comparison to the partially data-driven MNE-AR model is used to establish a baseline for current simulator performance, while comparison with the Jacobian-free model is used as an ablation study to investigate the impact of the Jacobian on preserving spatiotemporal information across channels.}

\subsubsection{Autoregressive Simulator}

As a baseline comparison, we use the AR fNIRS simulator \citep{santosa2018nirs} implemented in the MNE-NIRS toolbox \citep{gramfort2013meg, luke2021analysis}. This simulator generates synthetic optical density data by combining task-evoked hemodynamic responses with temporally autocorrelated noise. To match the experimental recordings, we calculate the cross-channel covariance matrix from the experimental data and scale the standard deviations of the simulated optical densities to match those of the experimental recordings.

\subsubsection{Jacobian-Free Ablated Simulator}
To test the claim that the sensitivity matrix is necessary to adequately represent the spatial relationships among local hemodynamic responses, we compare our simulator with an ablated version that does not rely on the Jacobian $\mathbf{J}$ to generate synthetic fNIRS signals. 

\review{Instead of projecting mesh-level absorption changes spatially via the Jacobian (\autoref{eq:rytov}), we construct effective channel-level absorption changes directly from the regional cortical responses and systemic physiology.
We introduce a region-to-channel mapping matrix $\mathbf{D}_C\in\mathbb{R}^{L\times C}$, initialized with binary entries indicating the assignment of cortical regions to measurement channels.
To simulate channel crosstalk, uniform noise ranging from $0$ to $0.1$ is added to the matrix, and each column is normalized to sum to $1$.
Using the wavelength-dependent cortical scaling factors $\rv{s}_C(\lambda;\rvtheta_C)$, the Jacobian-free simulator combines cortical and systemic absorption contributions analogously to \autoref{eq:deltamua}:
\begin{equation}
    \rvdelta\rvmu_{a,\mathrm{JF}}(\lambda,t;\rvtheta)
    = g_1\,\mathbf{D}_C^\top\left(\rv{s}_C(\lambda;\rvtheta_C)\odot\rvc(t;\rvtheta_C)\right)
    + g_2\,\,s_P(\lambda;\rvtheta_P)\,\rvdelta\tilde\rvmu_{P}(t;\rvtheta_P).
\end{equation}
Here, $\rvdelta\rvmu_{a,\mathrm{JF}}(\lambda,t;\rvtheta)\in\mathbb{R}^C$ denotes the effective channel-level absorption changes, and $\rvdelta\tilde\rvmu_{P}(t;\rvtheta_P)\in\mathbb{R}^C$ represents the globally shared physiological waveform replicated across the $C$ measurement channels.
The factor $s_P(\lambda;\rvtheta_P)$ provides the physiological HbO/HbR wavelength scaling.
The heuristic scaling factors are set to $g_1=500$ and $g_2=1500$ to roughly match the scale of the proposed simulator.
External artifact generation is identical to our full simulator.}

\subsection{Validation}\label{sec:Metrics}

\subsubsection{Model Ablation}\label{sec:Ablation}

To assess the contribution of each model component described in \autoref{sec:Model}, we perform an additive ablation study. 
Starting from the cortical activity model alone, we progressively add systemic physiology, external artifacts, and the full set of model components. 
\review{Specifically, we generate synthetic finger-tapping data under four conditions using fixed random seeds to ensure identical realizations of shared model components:} (1) systemic physiology only, (2) cortical activity only, (3) cortical activity with systemic physiology, (4) cortical activity with systemic physiology and external artifacts, which is the full model.

We then examine how each added component contributes to the cross-channel spatiotemporal structure of the simulated measurements by comparison with the real measurements.
For each condition, we estimate the correlation matrix across selected long-separation channels located over the region of interest (ROI), here the right primary motor cortex (M1) during a left-hand finger-tapping task. 
To illustrate the spatially localized effect of the HRF, we also include a distant channel over the left somatosensory cortex. Correlation matrices are computed for each trial, averaged across trials, and compared across ablation conditions.

\subsubsection{Channel Correlation Analysis}
To assess how well temporal features are preserved across channels, we calculated the cross-channel correlation matrix of the $\Delta$HbO signal for the long-separation channels above the ROI. 
We then compared the correlation matrices, averaged across trials, to obtain a single average correlation matrix for each simulated and experimental dataset.
Using the structural similarity index metric (SSIM), we compared how well each simulator preserves spatiotemporal information across the average cross-channel correlation matrix. Because the finger tapping task involves left-handed finger tapping, we selected channels above the right M1 region of the cortex as the ROI. For the bimanual surgical training task used for the surgical skill assessment, we designated channels above the left and right M1 as the ROI.

\subsubsection{Spectral Analysis}

Finally, we compare the systemic physiology in the experimental and synthetic data using spectral features from short-separation channels. Optical density data are first standardized independently for each channel. 
We then compute the continuous wavelet transform (CWT) for all short-separation channels and average the spectra across channels and trials.
The CWT is computed using a complex Morlet wavelet with a center frequency of 1.5 and a bandwidth of 1.0.
As in \cite{reddy2021evaluation}, we use the time-averaged CWT as the spectral summary for each condition. Similarity between experimental and synthetic data is quantified using the coefficient of determination ($R^2$) between their time-averaged CWT spectra.

We omit a separate comparison for the Jacobian-free model because systemic physiology is generated identically in the Jacobian-free and full models. In the surgical skill assessment dataset, trial durations vary; therefore, each trial is cropped to the first 30 seconds for spectral analysis.
\section{Results}\label{sec:Results}
The following sections demonstrate the ability of our model to generate high-quality synthetic data with realistic local and global structure.

\begin{figure}[t]
    \centering
    \includegraphics[width=1.0\textwidth]{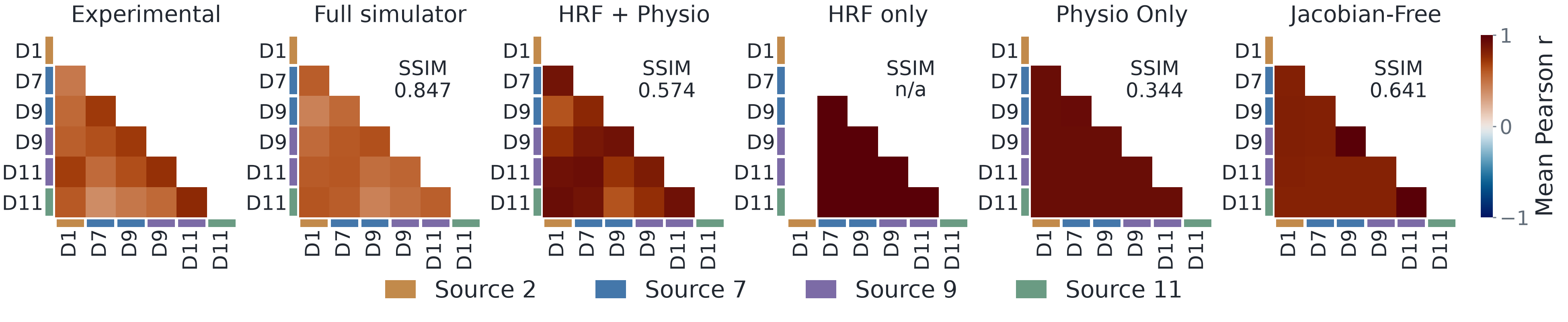}
    \caption{Average correlation matrices of $\Delta$HbO for experimental finger tapping and component-ablated simulated data. Channels are color coded based on the source (see \autoref{fig:corr_surgical} for the montage). Their locations are above the activated right primary motor cortex (\textcolor{brainblue}{S7-channels}, \textcolor{brainpurple}{S9-channels}), on the edge of it (\textcolor{braingreen}{S11-channel}) and distant on the left side of the head (\textcolor{brainyellow}{S2-channel}). Similarity between the experimental and simulated data is quantified using the structural similarity index measure (SSIM). Noteworthy, S2-D1 does not receive any signal, and thus does not contain variations, leading to mathematically undefined correlation values which is reflected in the assigned SSIM.}
    \label{fig:corr_tapping}
\end{figure}

\subsection{Ablation Study}

The ablation results follow the expected component-wise behavior (\autoref{fig:corr_tapping}). 
The full simulator best reproduces the heterogeneous cross-channel structure observed in the experimental data, with stronger correlations among channels that share either a source or detector and lower, less uniform correlations elsewhere. 
When external artifacts are removed, this behavior becomes more apparent, but correlations uniformly increase to unrealistically high values, indicating that these components are important for breaking the otherwise idealized channel structure.
The systemic physiology component produces broadly shared fluctuations and therefore induces high correlations across channels, consistent with its global nature. 
In contrast, the HRF-only condition induces correlations mainly among channels above the activated right M1 region, while the distant channel remains entirely uncorrelated (see, e.g., correlations with S2-D1). This distant channel receives no signal, and thus does not contain any variations, leading to mathematically undefined correlation values which is reflected in the assigned SSIM.
An overview of channel locations and the activated region for this finger tapping setting is shown in \autoref{fig:corr_surgical} (\textit{top-left}).

Channel S11-D11, near the edge of the ROI, receives a relatively weaker local HRF contribution than channels closer to the center of the ROI.
This reduced HRF amplitude does not necessarily appear as a lower correlation in the HRF-only condition, since correlation is insensitive to absolute signal scale.
However, in the component mixture (\autoref{eq:deltamua}), the weaker local HRF contribution increases the relative influence of the global systemic component, consistent with the elevated correlation between S11-D11 and the distant S2-D1 channel in the ``HRF + Physio'' panel and lower correlations with ROI-close channels S7-D9 and S9-D9.

The Jacobian-free model shows uniformly high correlations between channels, with higher correlation among channels that share detectors. Hence, without the Jacobian, the spatial encoding of absorption changes is missing, leading to unrealistic cross-channel structure.

Together, these trends show that the ablation isolates the expected spatial roles of the simulator components: localized cortical hemodynamics, global systemic physiology, and artifact variability that makes the full simulation more similar to experimental measurements.

\subsection{Cross-Channel Correlations}
\begin{figure}[t]
    \centering
    \includegraphics[width=1.0\textwidth]{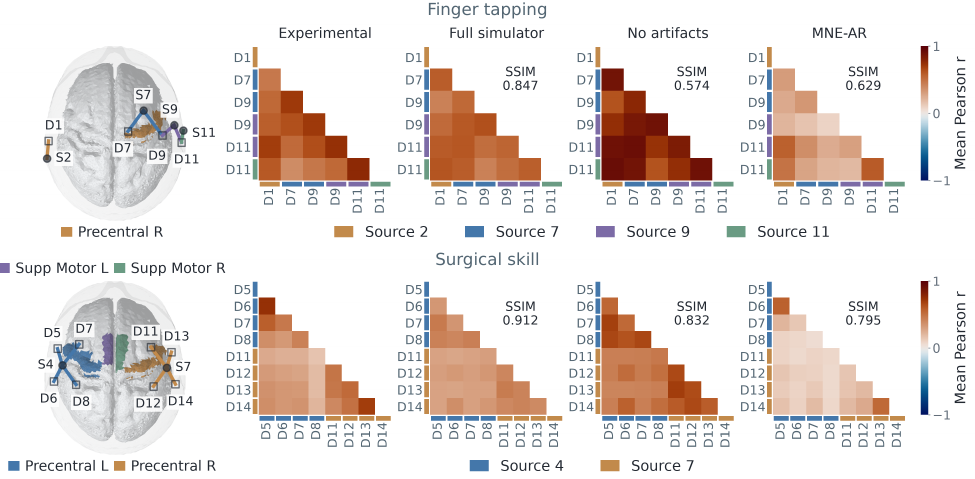}
    \caption{Similarity of averaged correlations between experimental $\Delta$HbO recordings and corresponding simulations across three models: the proposed model-based simulator with and without artifacts and the data-driven MNE-AR simulator. Correlations are shown for finger tapping (\textit{top}) and surgical skill (\textit{bottom}) settings, and similarity between the results is quantified using the structural similarity index measure (SSIM).
    The head mesh with active regions (\textit{left}) highlights source-detector pairings corresponding to the channels used for computing the correlation matrices. We selected active regions based on task relevance, and chose channels based on their proximity to these regions. For the finger tapping setting, a complementary channel is added (see \autoref{fig:corr_tapping}).}
    \label{fig:corr_surgical}
\end{figure}

\paragraph{Finger Tapping}
For the finger-tapping dataset, the experimental $\Delta$HbO correlation matrix over right M1 shows the expected spatial structure (\autoref{fig:corr_surgical}, \textit{top}). 
Channels that are spatially close (e.g., channels S7-D9 and S9-D9) tend to exhibit stronger correlations than more separated channel pairs (e.g., channels S7-D9 and S11-D11).
Our full simulator preserves this local correlation structure, with higher correlations concentrated among neighboring channels rather than distributed uniformly across the ROI.
In contrast, simulation without artifacts produces an overly homogeneous correlation pattern, reducing the distinction between local and distant channel pairs. 
\review{The MNE-AR model does not show a similar trend of increased correlation between neighboring channels; instead, we observe consistently lower correlations than those of experimental data, with no clear pattern between channel correlations and channel distance or proximity to the ROI.}

\paragraph{Surgical-skill assessment} For the surgical-skill assessment dataset, the experimental $\Delta$HbO correlation matrix shows a similar spatial organization across channels over the left and right M1 regions (\autoref{fig:corr_surgical} \textit{bottom}). 
Channels within the same local motor region tend to be more strongly correlated, whereas correlations between more distant channel groups across motor regions are weaker. 
The full simulator preserves this block-like structure, indicating that the sensitivity-matrix-based measurement model captures the spatial localization of the underlying hemodynamic responses. 
As observed with the finger tapping dataset, removing artifacts produces correlations that are too broadly elevated across channels, reinforcing the role of artifacts. 
The MNE-AR model again shows consistently lower correlations than the experimental data, however high correlations can be observed between a pair of channels in the left M1 (S4-D5 and S4-D6) and a pair of channels in the right M1 (S7-D13 and S7-D14), which are not observed in any other neighboring channels.
Across both datasets, these results indicate that the full simulator preserves localized hemodynamic structure and faithfully decorrelates signals, whereas the MNE-AR baseline fails to reproduce the spatially graded empirical cross-channel correlations.

\subsection{Continuous Wavelet Transform}
\begin{figure}[t]
    \centering
    \includegraphics[width=1.0\textwidth]{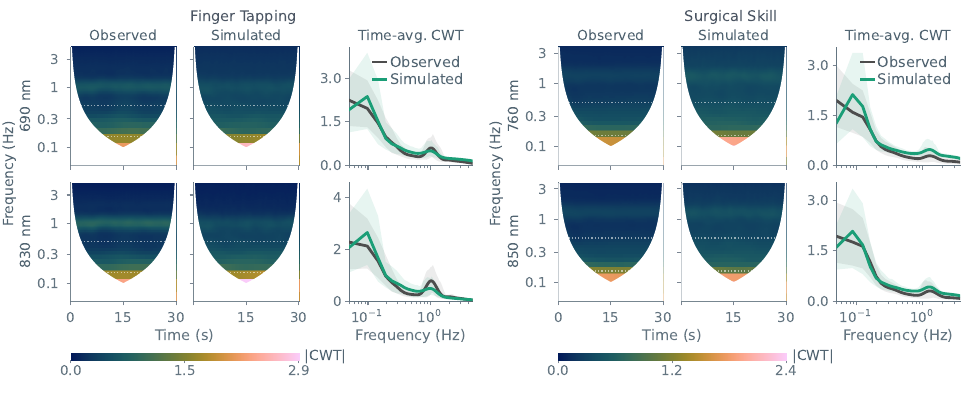}
    \caption{Continuous wavelet transforms (CWTs) and their time-averaged profiles for short-separation channels in the finger tapping (\textit{left}) and surgical skill assessment (\textit{right}) datasets. The simulated spectra closely follow the experimentally observed spectra, indicating that the simulator reproduces the dominant physiological frequency structure.}
    \label{fig:CWT}
\end{figure}

We further compare the spectral structure of the simulated and experimental short-separation channels using the continuous wavelet transform (CWT; \autoref{fig:CWT}). 
Across both datasets, the simulated CWTs reproduce the dominant frequency structure in the empirical data, with pronounced activity in the low-frequency and cardiac bands and visible contributions in the respiratory range.
This agreement is also reflected in the time-averaged CWT spectra, where the simulated and observed curves show almost identical peak locations and overall similar spectral profiles.

For the finger-tapping dataset, the time-averaged simulated and experimental spectra show strong agreement across both wavelengths, with $R^2$ estimates of $0.93$ and $0.91$ for 690 and 830 nm, respectively. 
Notably, the relative power of cardiac activity is higher in the smaller wavelength $\lambda_1$. This difference is not evident in the simulated data, suggesting that the relative change in HbO to HbR in systemic physiology $\rv\xi$ could be further tuned to more closely align simulated outcomes with the experimental data.
Similarly high agreement in time-averaged CWT is observed for the surgical-skill assessment dataset, with the exception of a mismatch at the lowest frequency bins, resulting in lower $R^2$ estimates of $0.78$ and $0.90$ for 760 nm and 850 nm, respectively.

\subsection{Global Summary Statistics}

\begin{figure}[t]
    \centering
    \includegraphics[width=1.0\textwidth]{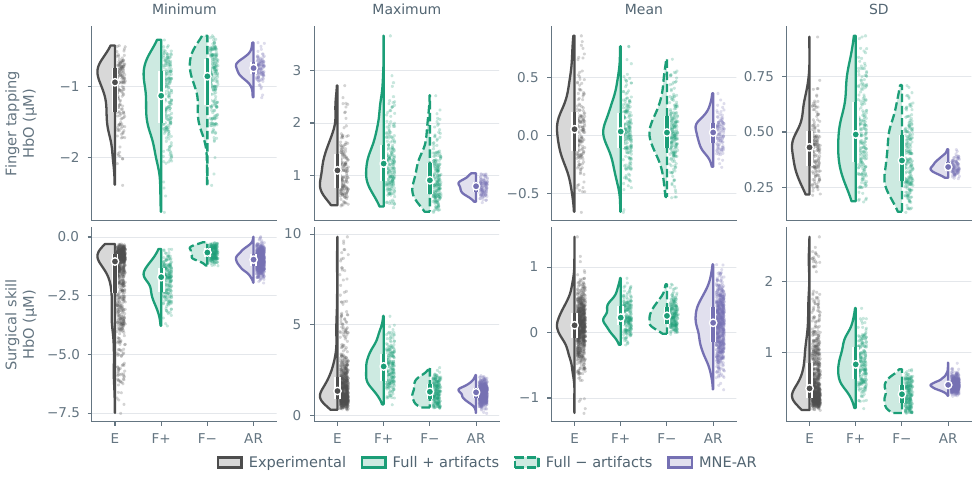}
    \caption{
    Primary global summary statistics for experimental and simulated fNIRS signals in the finger-tapping and surgical-skill assessment datasets. Violin plots compare the distributions of minimum, maximum, mean, and standard deviation across the experimental data, the full simulator with and without external artifacts, and the MNE-AR baseline.}
    \label{fig:summaryStats}
\end{figure}

Next, we compare global summary statistics of the experimental and simulated signals (\autoref{fig:summaryStats}; additional statistics in \autoref{fig:summaryStatsAppendix} and \autoref{fig:pain_summaryStats}). We consider the first four moments, minimum, maximum, range, and energy to assess whether the simulated signals fall within a plausible empirical range of signal scale, variability, and tail behavior. These descriptors are commonly used to characterize fNIRS responses across neurophotonics applications \citep{eser2025decoding, lee2024functional, wang2025detecting, eastmond2022deep}.

For the finger-tapping dataset, the full simulator with external artifacts provides the most consistent coverage of the empirical summary-statistic distributions. 
The means are well centered, and the extrema, standard deviations, range, energy, skewness, and kurtosis fall within a broadly comparable regime (\autoref{fig:summaryStats}; \autoref{fig:summaryStatsAppendix}). 
Removing external artifacts narrows the simulated distributions, especially for extrema, standard deviation, range, and energy. 
This supports the role of external artifacts in the full simulator in broadening signal variability, consistent with the correlation-based ablation results in \autoref{fig:corr_tapping}. 
The MNE-AR baseline also captures aspects of the signal scale, as expected for a model fitted directly to experimental data, but it exhibits less consistent behavior across statistics, including narrower distributions across most statistics except mean.

For the surgical-skill assessment dataset, using the parameter settings summarized in \autoref{tab:surgical_parameters}, our parameterization captures several central aspects of the empirical distributions, particularly the mean shift, but misses the larger extreme values present in the experimental data (\autoref{fig:summaryStats}).
Overall, these results show that the proposed simulator produces summary statistics in a realistic empirical range, while also confirming that global moments provide only a coarse validation of the simulated signal structure.

\section{Discussion}\label{sec:discussion}
Large-scale labeled fNIRS datasets are difficult to obtain, as experimental recordings are constrained by subject recruitment, task design, and measurement variability. 
This motivates the development of synthetic data generators that can produce controlled signals with known underlying sources and fully configurable nuisance components. 
In this paper, we introduced a simulation toolkit that combines the gold standard for modeling light transport through biological media \citep{fang2019graphics} with a theory-driven model intended to capture known fNIRS signal contributions, including artifacts \citep{brigadoi2014motion,gao2022deep}, systemic physiology \citep{hocke2018automated,nguyen2018adaptive}, and hemodynamic responses to cortical activity \citep{lindquist2009modeling,glover1999deconvolution}. 
By leveraging the Rytov approximation \citep{oleary1996imaging}, we can efficiently and accurately map simulated spatiotemporal changes in hemoglobin concentration to synthetic intensity measurements and their derivatives across the experimental probe (see~\autoref{sec:appdx-simulator-details}).

Across our initial validation experiments, the proposed simulator captures several complementary properties of empirical fNIRS recordings.
The global summary statistics show that simulated signals occupy a plausible range of amplitudes and variability for short recordings, while the CWT analysis demonstrates that most dominant physiological frequency components are reproduced. 
At the same time, the cross-channel correlation analyses indicate that the simulator preserves spatially structured hemodynamic information across the probe. 
Thus, the main strength of the proposed model is its ability to jointly reproduce global signal scale, physiological spectral structure, and localized channel relationships.

The comparison with a Jacobian-free simulator highlights the importance of our sensitivity-aware measurement model. 
The Jacobian maps localized cortical absorption changes through the head volume to probe-specific intensity measurements, preserving how cortical activity is expressed across nearby and distant channels. 
Without this sensitivity mapping, resulting channel correlations are not tied to the spatial organization of the probe and the relative position of the cortical activation, as observed in experimental recordings (\autoref{fig:corr_surgical}).

The component ablation further shows the additive contribution of the simulator's main components in physiologically interpretable ways. 
Systemic physiology introduces fluctuations shared across channels, whereas cortical hemodynamic responses remain localized around the activated region. 
External artifacts increase signal variability and widen the simulated distributions, improving agreement with several empirical summary statistics.
Taken together, these comparisons demonstrate the fidelity and mechanistic control of our approach.

The MNE-AR baseline captures selected statistical features, as expected for a statistical model estimated directly from experimental recordings, but it does not provide explicit control over systemic physiology and external artifacts, or spatial localization of cortical activations.
In contrast, our simulator combines a physics-based measurement model with interpretable signal components, making it suitable for generating synthetic fNIRS data with and without physiological contributions.

\review{An additional strength of our framework is its modularity, which facilitates integration with complementary neuroimaging modalities. For instance, although not explored in the present work, correlated regional activity could be introduced through joint priors over HRF parameters without modifying the optical forward model. Moreover, functional MRI data could inform the spatial activation maps in $\mathbf{A}$ and potentially guide the specification of dependencies between regional responses \citep{huppert2008direct,delaire2025nirstorm}.}

\paragraph{Limitations}
\review{Despite these promising results, our approach exhibits several limitations, which we plan to address in subsequent work. 
Some of these limitations are inherent to the nature of theory-driven approaches.
Accurately reproducing experimental recordings requires informative prior distributions for physiology, noise, and artifacts, which vary with optode placement, device properties, and experimental conditions. While overly broad distributions may generate implausible signals, dataset-specific calibration requires time-consuming iterative adjustments. Future work could automate this process through agent-guided prior-predictive calibration, refining parameter distributions based on discrepancies between simulated and experimental recordings.}

\review{Longer recordings and extended task periods reveal additional limitations in the modeling of physiological and hemodynamic responses. For sufficiently long recordings, such as those in the surgical skill assessment setting, we observed a persistent underestimation of low-frequency power (see \autoref{fig:CWT}) and discrepancies in summary statistics (see \autoref{fig:summaryStats}). Incorporating very-low-frequency oscillations \citep{hocke2018automated, bicciato2021increase} into the periodic physiology module may address these discrepancies.
Furthermore, the double-gamma HRF model has known limitations in representing responses to prolonged stimulation \citep{lindquist2009modeling}, potentially restricting its suitability for extended cognitive tasks. The modular simulator architecture allows alternative, more flexible HRF models, such as Gaussian processes \citep{eickenberg2017gaussian}, to be explored.}

\review{Although motion artifacts arise from changes in optode-scalp coupling, our current implementation samples them independently across measurement channels, neglecting potential dependencies between channels sharing an optode. Future extensions could model motion events at the optode level and propagate their effects to the corresponding channels, with an additional dependency structure informed by experimental recordings.}

\review{Moreover, task-evoked systemic physiological responses can confound fNIRS measurements \citep{kirilina2012physiological}. However, the current physiology module generates fluctuations from a time-invariant PSD, limiting its ability to capture such nonstationary responses. Future extensions could incorporate stimulus-dependent spectral modulation or oscillator-based models to simulate task-evoked physiological dynamics and broaden the range of experimental conditions that can be emulated.}

\review{Lastly, the first-order Rytov approximation is well established in diffuse optical imaging \citep{arridge1999optical,yao2018direct,wheelock2019high,guan2025continuous} and has demonstrated advantages over the first-order Born approximation across a range of tested absorption perturbations \citep{sassaroli2024revisiting}. Nevertheless, it can introduce errors in simulated intensities, with accuracy depending on the magnitude and spatial distribution of absorption changes.}


\review{Despite these limitations, the mechanistic and modular design of our simulator allows continuous refinements to improve fidelity and broaden the range of experimental conditions that can be modeled and simulated.}
\section{Conclusion}
We presented a spatiotemporally accurate, physiologically informed, full-head fNIRS simulator that combines mesh-based Monte Carlo photon transport with biophysically plausible models of cortical activation, systemic physiology, and external artifacts. 
By leveraging anatomically derived sensitivity matrices and the Rytov approximation, our simulator preserves probe-specific and wavelength-dependent spatiotemporal fidelity.
Across different experimental paradigms, it reproduced key properties of empirical fNIRS data, including physiological spectral peaks, cross-channel HRF correlation structure, and global signal statistics.
Looking ahead, access to pairs of simulated recordings and underlying model parameters could enable simulation-based inference for dataset-specific calibration without manual prior tuning.

Incorporating more flexible HRF models and subject-specific anatomical meshes would further improve realism in complex paradigms, while inclusion of task-dependent physiology \review{would introduce more realistic task-dependent confounders}.
Finally, extending the framework to multimodal settings, such as combined fNIRS–EEG or stimulation studies, could broaden its utility as a general-purpose toolkit for methodological validation and translational neuroimaging research.
\section{Acknowledgments}
Research reported in this publication was supported by the National Institute of Biomedical Imaging and Bioengineering of the National Institutes of Health under Award Numbers F31EB035931 and R01EB037715. The content is solely the responsibility of the authors and does not necessarily represent the official views of the National Institutes of Health. 
\bibliography{references}

\appendix
\appendix
\renewcommand{\thetable}{A\arabic{table}}
\setcounter{table}{0}
\renewcommand{\thefigure}{A\arabic{figure}} 
\setcounter{figure}{0} 
\section{Additional results}

\subsection{Pain Assessment Experiment}

\subsubsection{Experimental setting} In addition to the finger tapping recordings, the dataset described in \cite{yucel2015short} includes an electrical stimulus pain assessment experiment that uses 11 of the same subjects (with one subject excluded), fNIRS spectrometer, and optode montage. This study used a 5 Hz electrical stimulator (Neurometer CPT, Neurotron, Baltimore, Maryland), which was attached to the left thumb of each subject.
Prior to the start of the recording, subjects had the electrical stimulator calibrated to determine subjective pain ratings of 3/10 (innocuous) and 7/10 (noxious). After recording the finger tapping task described in \autoref{sec:experimental-data}, subjects underwent the pain assessment task, with 5 s of electrical stimulus followed by 25 s of rest. Each recording consisted of 12 noxious and 12 innocuous stimuli, randomly ordered, for a total of 12.5 minutes per run. Due to one recording stopping early, we have a total of 123 noxious and 129 innocuous trials.

\subsubsection{Pain assessment results}\label{sec:pain-assessment-results}
\begin{figure}[t]
    \centering
    \includegraphics[width=1\textwidth]{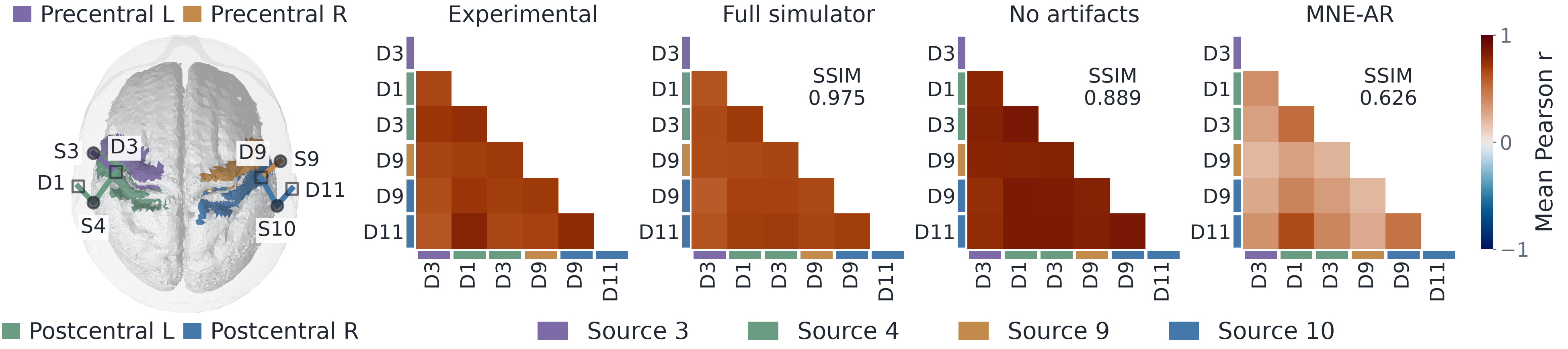}
    \caption{Mean cross-channel correlations of experimental and simulated $\Delta$HbO recordings for the pain assessment task. Correlation matrices are compared across the simulator with and without artifacts, and MNE-AR, with similarity quantified by SSIM. The head mesh (\textit{left}) highlights bilateral precentral and postcentral regions and the corresponding source-detector pairings used in the analysis \autoref{sec:pain-assessment-results}.
    }
    \label{fig:Pain_Correlation_Matrix}
\end{figure}

\paragraph{Correlation analysis}
In the analysis of the pain assessment dataset, we selected the previously identified channels of interest \citep{yucel2015specificity}, covering parts of the left and right somatosensory regions and motor cortex. In the mean correlation matrices \autoref{fig:Pain_Correlation_Matrix}, we observed uniformly high correlation and, unlike the surgical skill assessment dataset, did not observe a decrease in correlation between channels on the left and right sides of the head. This behavior is mimicked in the full simulator, and as seen in previous datasets \autoref{fig:corr_surgical}, removing artifacts increases correlation to unrealistically high values. We continue to observe low correlations in the MNE-AR model. Compared to the finger tapping and surgical skill assessment datasets, clear spatial trends in correlation were not obvious in this dataset, possibly due to a less localized response to the painful stimuli or the increase in task-evoked global physiological contributions.

\begin{figure}[t]
    \centering
    \includegraphics[width=0.5\textwidth]{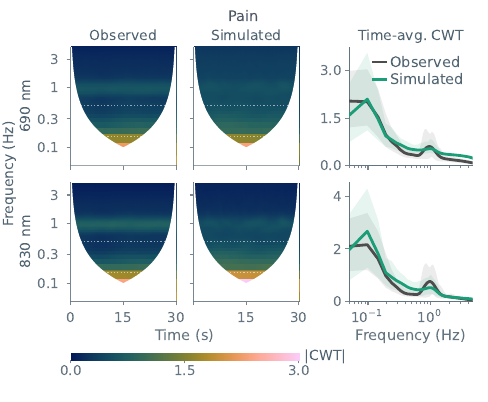}
    \caption{Continuous wavelet transform (CWT) of the short separation channels between experimental pain assessment and simulated data, along with the corresponding time-averaged CWT.}
    \label{fig:Pain_CWT}
\end{figure}

\paragraph{Spectral analysis}
As in \autoref{sec:Results}, we see clear low-frequency peaks and cardiac peaks in both the experimentally observed and simulated pain assessment data. As in the surgical skill assessment dataset, our simulator produces too little power at the lowest frequency bins for one wavelength, resulting in $R^2$ scores of $0.78$ and $0.90$ for the 690 and 830 nm wavelengths, respectively (\autoref{fig:Pain_CWT}).

\begin{figure}[t]
    \centering
    \includegraphics[width=0.99\linewidth]{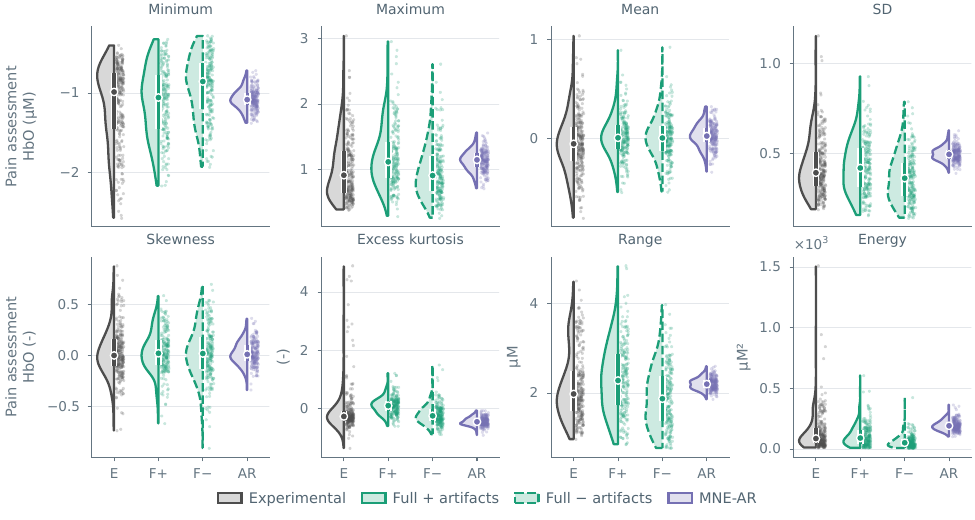}
    \caption{Global summary statistics for experimental pain assessment fNIRS signals compared with the full simulator with and without external artifacts, and the MNE-AR baseline.}
    \label{fig:pain_summaryStats}
\end{figure}

\paragraph{Summary statistics}
When comparing the summary statistics for the experimental pain assessment dataset with those of the benchmark models (\autoref{fig:pain_summaryStats}), there is observable underdispersion in kurtosis in the simulated data with and without motion artifacts. 
In addition, the proposed simulator with motion artifacts shows comparable, though not identical, distributions across most other summary statistics, while MNE-AR remains significantly underdispersed.

\subsection{Summary Statistics}
\begin{figure}[t]
    \centering
    \includegraphics[width=0.99\linewidth]{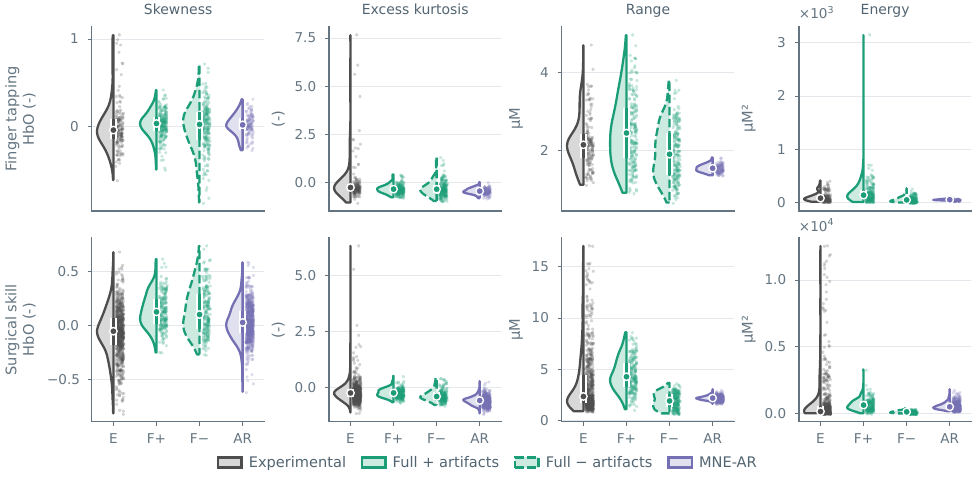}
    \caption{
    Additional secondary global summary statistics for experimental and simulated fNIRS signals in the finger-tapping and surgical-skill assessment datasets. Violin plots show skewness, kurtosis, range, and signal energy. These statistics are more sensitive to tail behavior and rare high-amplitude events than the first-order statistics shown in \autoref{fig:summaryStats}.
    }
    \label{fig:summaryStatsAppendix}
\end{figure}

In addition to the statistics provided in \autoref{fig:summaryStats}, we present four additional statistics shown in \autoref{fig:summaryStatsAppendix}, commonly used for analyzing fNIRS data  \citep{eser2025decoding, lee2024functional, wang2025detecting, eastmond2022deep}. Summary statistics for the pain assessment task are shown in \autoref{fig:pain_summaryStats} and are discussed in the appendix section above.

\begin{figure}[t]
    \centering
    \includegraphics[width=0.99\linewidth]{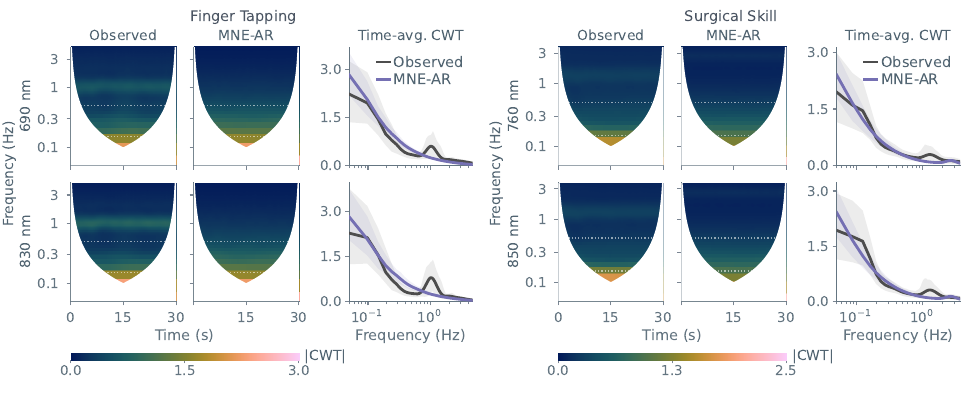}
    \caption{Continuous wavelet transform (CWT) of the short separation channels from the finger tapping (\textit{left}) and surgical skill assessment (\textit{right}) settings from experimental recordings and the data-driven MNE-AR modeling approach.}
    \label{fig:MNE_CWT}
\end{figure}

\subsection{MNE-AR CWT}
Here, we present the CWT for the MNE-AR model and compare it with the experimental finger-tapping and surgical skill assessment datasets (\autoref{fig:MNE_CWT}). Unlike our proposed model, the CWT of the MNE-AR model does not capture the cardiac peak present in either experimental dataset; instead, it broadly follows a $1/f$ exponential curve.

\subsection{FOOOF Spectral Analysis}
Because the CWT plots show an aggregated average of the trials' spectra, we use the spectral parameterization tool from the FOOOF toolbox \citep{donoghue2020parameterizing} to estimate the systemic physiology peaks in the PSD of each recording. 
Here we take an example short separation channel from each simulated and experimental trial, and plot the peaks detected by the FOOOF algorithm in each frequency band of interest. We plot the distributions of simulated data from our full simulator with artifacts and the MNE-AR model as ellipses centered on the mean peak within each band, with a radius corresponding to three standard deviations of the data. We then plot the estimated peaks from the experimental data and calculate how many of the experimental peaks fall within the ellipse for each simulator.

\subsubsection{FOOOF Results}

\begin{table}
    \centering
    \begin{tabular}{@{}clcccccc@{}}
        \toprule
        \multirow{2}{*}{$\lambda$ (nm)}
        & \multirow{2}{*}{Physiological band}
        & \multicolumn{2}{c}{\textbf{Finger Tapping}}
        & \multicolumn{2}{c}{\textbf{Pain}}
        & \multicolumn{2}{c}{\textbf{Surgical Skill}} \\
        \cmidrule(lr){3-4} \cmidrule(lr){5-6} \cmidrule(lr){7-8}
        & & Ours & MNE-AR & Ours & MNE-AR & Ours & MNE-AR \\
        \midrule
        \multirow{3}{*}{690 / 760}
        & LFO         & 0.97 & 0.95 & 1.00 & 0.81 & 0.99 & 0.94 \\
        & Respiratory & 0.92 & 0.97 & 0.94 & 0.96 & 1.00 & 1.00 \\
        & Cardiac     & 0.63 & 0.47 & 0.75 & 0.46 & 0.87 & 0.49 \\
        \addlinespace[0.4em]
        \multirow{3}{*}{830 / 850}
        & LFO         & 0.95 & 0.95 & 0.90 & 0.94 & 0.98 & 0.90 \\
        & Respiratory & 0.90 & 0.93 & 0.96 & 0.97 & 1.00 & 1.00 \\
        & Cardiac     & 0.96 & 0.20 & 0.94 & 0.23 & 0.93 & 0.45 \\
        \bottomrule
    \end{tabular}
    \caption{Fraction of detected experimental systemic
    physiology peaks captured by the descriptive
    3-standard-deviation covariance ellipse of
    simulation peaks. Ours includes artifact simulation.
    Wavelengths are 690/830 nm for finger tapping and
    pain, and 760/850 nm for surgical skill; paired
    wavelength labels follow this order.}
    \label{tab:Physio_Results}
\end{table}

\begin{figure}[t]
    \centering
    \includegraphics[width=1.0\textwidth]{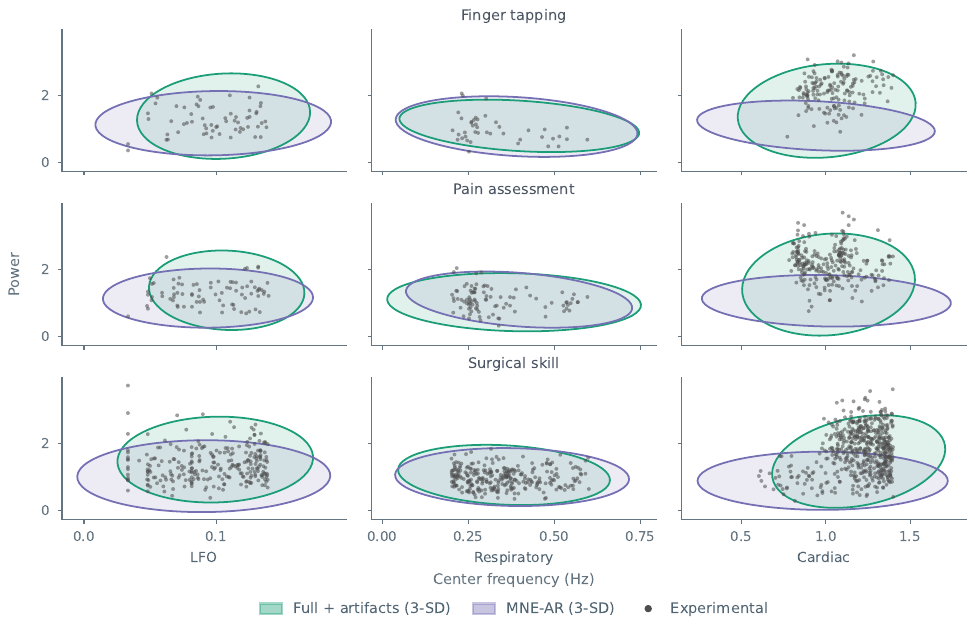}
    \caption{FOOOF-derived \citep{donoghue2020parameterizing} spectral peaks in the LFO, respiratory, and cardiac bands across finger tapping, pain assessment, and surgical skill datasets. Experimental peaks (gray points) are compared with descriptive 3-SD covariance ellipses for the full simulator with artifacts (teal) and MNE-AR (purple). Coordinates represent peak center frequency and peak height above the aperiodic fit ($\log_{10}$ power). We extracted peaks from 30-second optical-density recordings using one fixed short-separation channel per dataset, with 830 nm for finger tapping and pain assessment and 850 nm for surgical skill.}
    \label{fig:finger_tapping_FOOOF}
\end{figure}

Across datasets, our simulator captures the majority of the peaks across all frequency bands in the finger tapping dataset \autoref{tab:Physio_Results}. In contrast, the MNE-AR model performs comparably well across the LFO and respiratory frequency bands but greatly underperforms in the cardiac frequency bands. Notably, this is because the MNE-AR model consistently underestimates power at the cardiac-band peaks in the experimental data \autoref{fig:finger_tapping_FOOOF}, capturing a substantially smaller fraction of the cardiac peaks present in the finger tapping and pain assessment datasets.
In the pain assessment dataset, our full simulator performs worse in the LFO frequency band \autoref{tab:Physio_Results}, missing a few detected peaks at the extreme low end of the LFO frequency band \autoref{fig:finger_tapping_FOOOF}. This potentially indicates a missing very-low-frequency band \citep{hocke2018automated, lindquist2009modeling}, as mentioned in \autoref{sec:discussion}.


\section{Simulator details}\label{sec:appdx-simulator-details}
\subsection{Mesh-Based Monte Carlo Simulations}
Using the command-line interface of MMC V2025.10 (Bubble Tea), each optode is simulated via the simulation parameters listed in \autoref{tab:MMC_Inputs} and the relevant optical properties listed in \autoref{tab:optical-properties}. With the given input parameters, it takes  $\approx$5 hours on an NVIDIA GeForce RTX 4090 to simulate the entire Jacobian for both wavelengths of the larger finger tapping/pain assessment montage.

\begin{table}[ht]
    \centering
    \begin{tabular}{lcl}
        \toprule
        Parameter &  Value & Description\\
        \midrule
        \texttt{nphoton} & 4e9 & Number of simulated photons\\
        \texttt{tstart} & 0 & Simulation starting time\\
        \texttt{tend} & 5e-9 & Simulation ending time\\
        \texttt{tstep} & 5e-9 & Duration of simulation time step\\
        \texttt{method} & \texttt{elem} &  Ray-tracing method\\
        \texttt{issaveexit} & 1 & Option to save photon exit positions and directions\\
        \texttt{issavedet} & 1 & Option to save photon info at detectors\\
        \texttt{outputtype} & \texttt{flux} & Output quantity\\
        \bottomrule
    \end{tabular}
    \caption{Input parameters for MMC simulations.}
    \label{tab:MMC_Inputs}
\end{table}

\begin{table}[t]
\centering
\small
\begin{tabularx}{\linewidth}{@{}llccc>{\raggedright\arraybackslash}X@{}}
\toprule
$\lambda$ (nm) & Tissue
& $\mu_a$ [mm$^{-1}$]
& $\mu_s$ [mm$^{-1}$]
& $g$ & Reference \\
\midrule
\multirow{5}{*}{690 / 830}
& Scalp        & 0.0159/0.0191 & 7.2730/6.0000 & 0.89/0.89 & \cite{strangman2003factors} \\
& Skull        & 0.0101/0.0136 & 9.0910/7.8180 & 0.89/0.89 & \cite{strangman2003factors} \\
& CSF          & 0.0004/0.0026 & 0.0091/0.0091 & 0.89/0.89 & \cite{strangman2003factors}, \cite{lewis2024revisiting} \\
& Gray matter  & 0.0200/0.0300 & 8.0000/7.0000 & 0.89/0.90 & \cite{lewis2024revisiting} \\
& White matter & 0.0700/0.0900 & 40.000/33.000 & 0.85/0.87 & \cite{lewis2024revisiting} \\
\addlinespace[0.5em]
\multirow{5}{*}{760 / 850}
& Scalp        & 0.0170/0.0190 & 6.7273/5.8182 & 0.89/0.89 & \cite{bonilauri2023impact} \\
& Skull        & 0.0116/0.0139 & 8.5455/7.6364 & 0.89/0.89 & \cite{bonilauri2023impact} \\
& CSF          & 0.0021/0.0026 & 0.0091/0.0091 & 0.89/0.89 & \cite{strangman2003factors}, \cite{lewis2024revisiting} \\
& Gray matter  & 0.0184/0.0260 & 7.5512/6.8682 & 0.90/0.90 & \cite{yaroslavsky2002optical} \\
& White matter & 0.0822/0.1000 & 37.930/34.200 & 0.87/0.88 & \cite{yaroslavsky2002optical} \\
\bottomrule
\end{tabularx}
\caption{Optical properties used in the Monte Carlo simulations.
Paired values follow the wavelength order in the first column.
$\mu_a$ and $\mu_s$ denote absorption and scattering coefficients,
respectively, and $g$ is the anisotropy factor.
Reduced scattering coefficients were converted where necessary
using $\mu_s=\mu_s'/(1-g)$.
The refractive index was fixed at $n=1.37$.}
\label{tab:optical-properties}
\end{table}

\subsection{Simulation Settings for Finger Tapping and Pain Datasets}\label{FingerTapping_Params}
\begin{table}[t]
    \centering
    \small
    \begin{tabularx}{\linewidth}{@{}cll>{\raggedright\arraybackslash}X@{}}
        \toprule
        & \textbf{Parameter} & \textbf{Prior} & \textbf{Description} \\
        \midrule
        \multirow{7}{*}{$\rvtheta_C$}
        & $d^+_\ell$ & $\operatorname{TN}_{[0.5,1.5]}(1,0.2)$
        & Positive gamma scale (s). \\

        & $d^-_\ell$ & $\operatorname{TN}_{[0.5,1.5]}(1,0.2)$
        & Negative gamma scale (s). \\

        & $\tau_\ell$ & $\operatorname{TN}_{[2.5,8]}(5,1.25)$
        & Time from stimulus onset to positive gamma mode (s). \\

        & $\delta_\ell$ & $\operatorname{TN}_{[5,13]}(8,2)$
        & Delay between positive and negative gamma modes (s). \\

        & $\kappa_\ell$ & $\operatorname{TN}_{[6,50]}(30,20)$
        & Regional response peak amplitude after convolution ($\mu$M). \\

        & $a^-_\ell$ & $\operatorname{TN}_{[0.02,0.6]}(0.32,0.16)$
        & Negative gamma weight relative to the positive component. \\

        & $\xi_{C,\ell}$ & $\mathcal{U}(0.67,0.75)$
        & HbO weight; HbR weight is $-(1-\xi_{C,\ell})$. \\
        \bottomrule
    \end{tabularx}
    \caption{Cortical activation parameter priors shared across the three datasets, specified for each active region $\ell=1,\ldots,L$. $\operatorname{TN}_{[a,b]}(\mu,\sigma)$ denotes a normal distribution with underlying mean $\mu$ and standard deviation $\sigma$, truncated to $[a,b]$.}
    \label{tab:cortical_parameters}
\end{table}
\begin{table}[t]
    \centering
    \small
    \begin{tabularx}{\linewidth}{
        @{}cll>{\raggedright\arraybackslash}X@{}
    }
        \toprule
        & \textbf{Parameter} & \textbf{Prior / Value}
        & \textbf{Description} \\
        \midrule

        \multirow{11}{*}{$\rvtheta_P$}
        & $\eta$
        & $\mathcal{U}(0.01,0.1)$
        & Constant offset in the physiological PSD. \\

        & $\boldsymbol{\alpha}$
        & $\operatorname{Dir}(7,0.0625,1.5)$
        & Relative PSD peak weights (Mayer, respiratory, cardiac). \\

        & $\nu_1$
        & $\operatorname{TN}_{[0.06,0.14]}(0.1,0.025)$
        & Mayer-wave center frequency (Hz). \\

        & $\nu_2$
        & $\operatorname{TN}_{[0.15,0.5]}(0.3,0.05)$
        & Respiratory center frequency (Hz). \\

        & $\nu_3$
        & $\operatorname{TN}_{[0.6,1.8]}(1,0.15)$
        & Cardiac center frequency (Hz). \\

        & $\rho$
        & $\mathcal{N}(1.8,0.1)$
        & Aperiodic PSD exponent ($1/f^\rho$). \\

        & $w_{\mathrm{PSD}}$
        & $\mathcal{U}(0,1)$
        & Relative weighting of periodic and aperiodic PSD components. \\

        & $R_{\mathrm{phys}}$
        & $\operatorname{TN}_{[0.2,7]}(0.5,0.14)$
        & Target RMS scaling of the systemic $\Delta$HbT component. \\
        
        & $\xi_P$
        & $\mathcal{U}(0.75,0.975)$
        & Relative HbO/HbR concentration change in systemic physiology. \\

        & $f_\xi(690\,\mathrm{nm})$
        & $\operatorname{TN}_{[0,1.2]}(1,0.2)$
        & Noise-to-RMS ratio of the Rytov-normalized measurement. \\

        & $f_\xi(830\,\mathrm{nm})$
        & $\operatorname{TN}_{[0,0.9]}(0.8,0.03)$
        & Noise-to-RMS ratio of the Rytov-normalized measurement. \\

        \midrule

        \multirow{9}{*}{\textit{Artifacts}}
        & $P(t_s)$
        & $0.05$
        & Spike occurrence probability per second. \\

        & $A_s$
        & $\mathcal{U}(-2,2)$
        & Spike amplitude relative to signal standard deviation. \\

        & $P(t_b)$
        & $0.05$
        & Configured baseline-shift occurrence probability. \\

        & $A_b$
        & $\mathcal{U}(-2,2)$
        & Baseline-shift amplitude relative to signal standard deviation. \\

        & $\epsilon$
        & $125$
        & Minimum intensity after motion artifact addition. \\

        & $P(d)$
        & $\mathcal{U}(0,0.025)$
        & Channel dropout probability. \\

        & $A_d$
        & $\mathcal{U}(0.05,0.15)$
        & Dropout noise amplitude relative to signal standard deviation. \\

        & $\epsilon_d$
        & $20$
        & Minimum intensity after channel dropout. \\

        \bottomrule
    \end{tabularx}
    \caption{Systemic physiology and artifact parameters
    used to generate synthetic \emph{finger tapping} data.
    The components of $\boldsymbol{\alpha}$ correspond
    to Mayer waves, respiration, and cardiac activity.
    $\operatorname{TN}_{[a,b]}(\mu,\sigma)$ denotes a
    normal distribution with underlying mean $\mu$
    and standard deviation $\sigma$, truncated to
    $[a,b]$.}
    \label{tab:fingertapping_parameters}
\end{table}

\begin{table}[t]
    \centering
    \small
    \begin{tabularx}{\linewidth}{
        @{}cll>{\raggedright\arraybackslash}X@{}
    }
        \toprule
        & \textbf{Parameter} & \textbf{Prior / Value}
        & \textbf{Description} \\
        \midrule

        \multirow{11}{*}{$\rvtheta_P$}
        & $\eta$
        & $\mathcal{U}(0.01,0.1)$
        & Constant offset in the physiological PSD. \\

        & $\boldsymbol{\alpha}$
        & $\operatorname{Dir}(7,0.0625,1.5)$
        & Relative PSD peak weights (Mayer, respiratory, cardiac). \\

        & $\nu_1$
        & $\operatorname{TN}_{[0.06,0.14]}(0.1,0.025)$
        & Mayer-wave center frequency (Hz). \\

        & $\nu_2$
        & $\operatorname{TN}_{[0.15,0.5]}(0.3,0.05)$
        & Respiratory center frequency (Hz). \\

        & $\nu_3$
        & $\operatorname{TN}_{[0.6,1.8]}(1,0.15)$
        & Cardiac center frequency (Hz). \\

        & $\rho$
        & $\mathcal{N}(1.8,0.1)$
        & Aperiodic PSD exponent ($1/f^\rho$). \\

        & $w_{\mathrm{PSD}}$
        & $\mathcal{U}(0,1)$
        & Relative weighting of periodic and aperiodic PSD components. \\

        & $R_{\mathrm{phys}}$
        & $\operatorname{TN}_{[0.4,1.3]}(0.5,1.14)$
        & Target RMS scaling of the systemic $\Delta$HbT component. \\

        & $\xi_P$
        & $\mathcal{U}(0.65,0.975)$
        & Relative HbO/HbR concentration change in systemic physiology. \\

        & $f_\xi(690\,\mathrm{nm})$
        & $\operatorname{TN}_{[0.6,1.4]}(0.8,0.2)$
        & Noise-to-RMS ratio of the Rytov-normalized measurement. \\

        & $f_\xi(830\,\mathrm{nm})$
        & $\operatorname{TN}_{[0,0.3]}(0.2,0.1)$
        & Noise-to-RMS ratio of the Rytov-normalized measurement. \\

        \midrule

        \multirow{8}{*}{\textit{Artifacts}}
        & $P(t_s)$
        & $0.02$
        & Spike occurrence probability per second. \\

        & $A_s$
        & $\mathcal{U}(-1.5,1.5)$
        & Spike amplitude relative to signal standard deviation. \\

        & $P(t_b)$
        & $0.02$
        & Configured baseline-shift occurrence probability. \\

        & $A_b$
        & $\mathcal{U}(-1.5,1.5)$
        & Baseline-shift amplitude relative to signal standard deviation. \\

        & $\epsilon$
        & $125$
        & Minimum intensity after motion artifact addition. \\

        & $P(d)$
        & $\mathcal{U}(0,0.025)$
        & Channel dropout probability. \\

        & $A_d$
        & $\mathcal{U}(0.05,0.15)$
        & Dropout noise amplitude relative to signal standard deviation. \\

        & $\epsilon_d$
        & $20$
        & Minimum intensity after channel dropout. \\

        \bottomrule
    \end{tabularx}
    \caption{Systemic physiology and artifact parameters
    used to generate synthetic \emph{pain assessment} data.
    The components of $\boldsymbol{\alpha}$ correspond
    to Mayer waves, respiration, and cardiac activity.
    $\operatorname{TN}_{[a,b]}(\mu,\sigma)$ denotes a
    normal distribution with underlying mean $\mu$
    and standard deviation $\sigma$, truncated to
    $[a,b]$.}
    \label{tab:pain_parameters}
\end{table}

\begin{table}[t]
    \centering
    \small
    \begin{tabularx}{\linewidth}{
        @{}cll>{\raggedright\arraybackslash}X@{}
    }
        \toprule
        & \textbf{Parameter} & \textbf{Prior / Value}
        & \textbf{Description} \\
        \midrule

        \multirow{11}{*}{$\rvtheta_P$}
        & $\eta$
        & $\mathcal{U}(0.01,0.1)$
        & Constant offset in the physiological PSD. \\

        & $\boldsymbol{\alpha}$
        & $\operatorname{Dir}(3.5,0.03,0.75)$
        & Relative PSD peak weights (Mayer, respiratory, cardiac). \\

        & $\nu_1$
        & $\operatorname{TN}_{[0.06,0.14]}(0.1,0.02)$
        & Mayer-wave center frequency (Hz). \\

        & $\nu_2$
        & $\operatorname{TN}_{[0.26,0.5]}(0.37,0.15)$
        & Respiratory center frequency (Hz). \\

        & $\nu_3$
        & $\operatorname{TN}_{[1,1.6]}(1.3,0.2)$
        & Cardiac center frequency (Hz). \\

        & $\rho$
        & $\mathcal{N}(2,0.1)$
        & Aperiodic PSD exponent ($1/f^\rho$). \\

        & $w_{\mathrm{PSD}}$
        & $\mathcal{U}(0.38,1)$
        & Relative weighting of periodic and aperiodic PSD components. \\

        & $R_{\mathrm{phys}}$
        & $\operatorname{TN}_{[0.28,0.91]}(0.35,0.8)$
        & Target RMS scaling of the systemic $\Delta$HbT component.\\

        & $\xi_P$
        & $\mathcal{U}(0.65,0.98)$
        & Relative HbO/HbR concentration change in systemic physiology. \\

        & $f_\xi(760\,\mathrm{nm})$
        & $\operatorname{TN}_{[0.34,1.26]}(0.6,0.18)$
        & Noise-to-RMS ratio of the Rytov-normalized measurement. \\

        & $f_\xi(850\,\mathrm{nm})$
        & $\operatorname{TN}_{[0,0.6]}(0.4,0.15)$
        & Noise-to-RMS ratio of the Rytov-normalized measurement. \\

        \midrule

        \multirow{8}{*}{\textit{Artifacts}}
        & $P(t_s)$
        & $0.02$
        & Spike occurrence probability per second. \\

        & $A_s$
        & $\mathcal{U}(-2,2)$
        & Spike amplitude relative to signal standard deviation. \\

        & $P(t_b)$
        & $0.02$
        & Configured baseline-shift occurrence probability. \\

        & $A_b$
        & $\mathcal{U}(-2,2)$
        & Baseline-shift amplitude relative to signal standard deviation. \\

        & $\epsilon$
        & $20$
        & Minimum intensity after motion artifact addition. \\

        & $P(d)$
        & $\mathcal{U}(0.08,0.15)$
        & Channel dropout probability. \\

        & $A_d$
        & $\mathcal{U}(0.05,0.15)$
        & Dropout noise amplitude relative to signal standard deviation. \\

        & $\epsilon_d$
        & $15$
        & Minimum intensity after channel dropout. \\

        \bottomrule
    \end{tabularx}
    \caption{Systemic physiology and artifact parameters
    used to generate synthetic \emph{surgical-skill assessment}
    data. The components of $\boldsymbol{\alpha}$ correspond
    to Mayer waves, respiration, and cardiac activity.
    $\operatorname{TN}_{[a,b]}(\mu,\sigma)$ denotes a
    normal distribution with underlying mean $\mu$
    and standard deviation $\sigma$, truncated to
    $[a,b]$.}
    \label{tab:surgical_parameters}
\end{table}

\begin{figure}[t]
    \centering
    \includegraphics[width=0.99\linewidth]{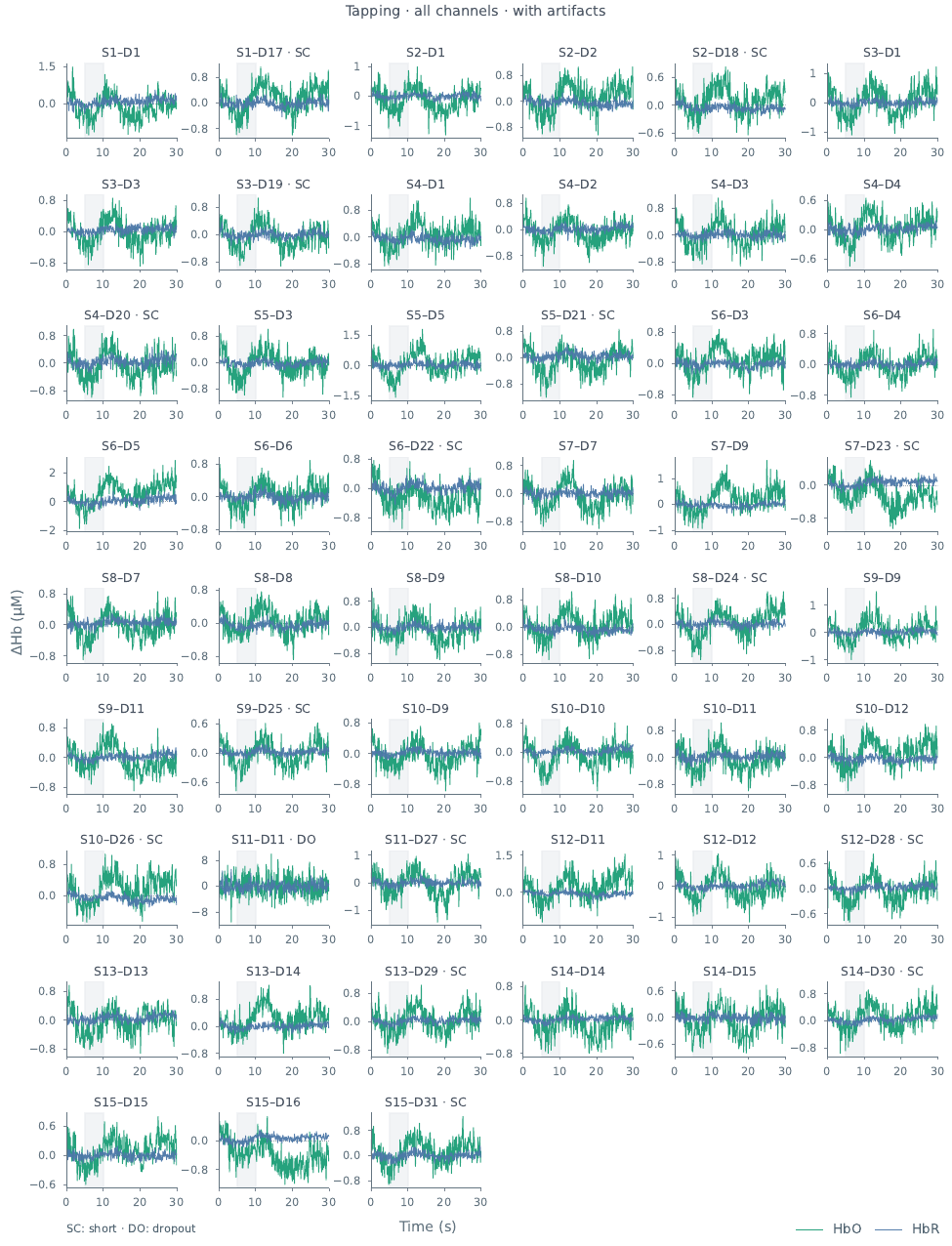}
    \caption{Simulated finger tapping data. Example of simulated chromophore changes for a multi-channel fNIRS device, with long- and short-separation channels corresponding to the finger tapping and pain assessment montage shown in \autoref{fig:Montage}.}
    \label{fig:FT_example1}
\end{figure}

Pain and finger-tapping data are simulated using 30-second trials with a stimulus onset of 5 seconds, a stimulus duration of 5 seconds, and a post-stimulus rest of 20 seconds, corresponding to the experimental protocol. 
All model hyperparameters used to simulate the finger-tapping and pain assessment datasets are listed in \autoref{tab:fingertapping_parameters} and \autoref{tab:pain_parameters}, respectively. Cortical parameters are the same throughout the experimental settings and are listed in \autoref{tab:cortical_parameters}.
An example simulated finger-tapping trial is shown in \autoref{fig:FT_example1}. In terms of computational efficiency, on an 11th Gen Intel Core i9-11900K, we simulate $1000$ 30-second finger-tapping trials in under ten minutes for Jacobian-informed simulations and less than one minute for Jacobian-free simulations. These simulation speeds do not include time for Jacobian generation. In comparison, the MNE-AR simulator takes under a minute to generate $100$ finger-tapping trials.

\subsection{Simulation Settings for the Surgical Skill Dataset}\label{Surgical_Params}
\begin{figure}[t]
    \centering
    \includegraphics[width=0.99\linewidth]{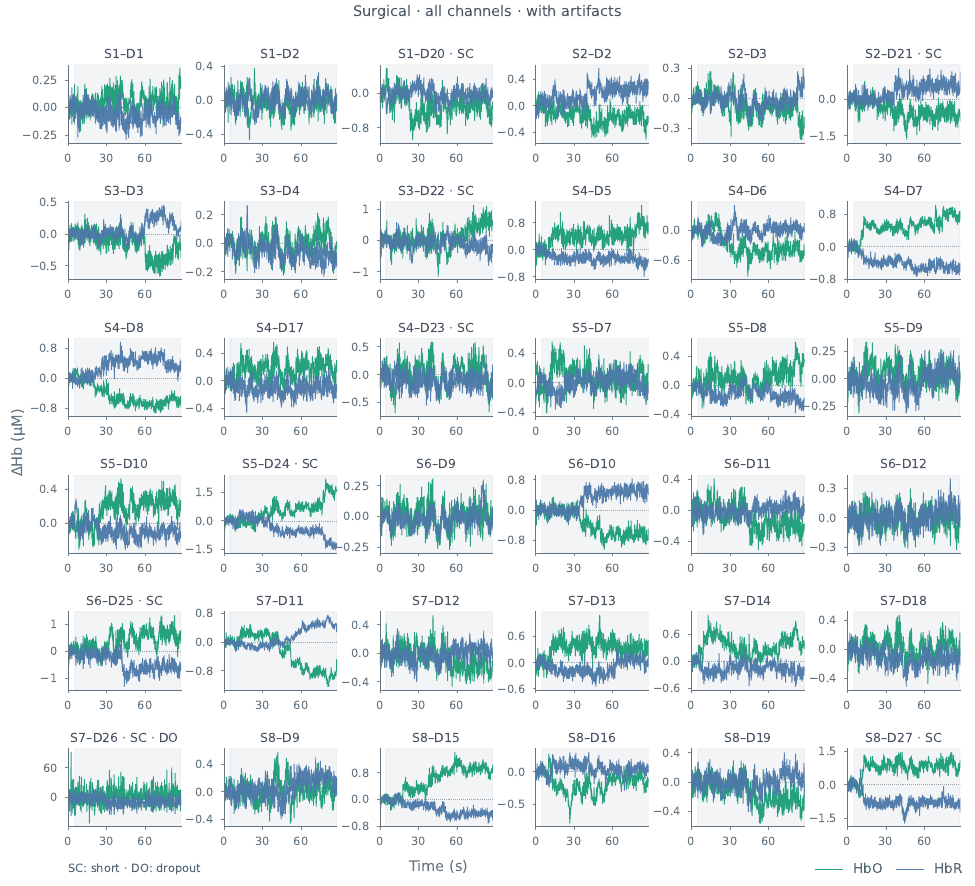}
    \caption{Simulated surgical skill assessment data. Example of simulated chromophore changes for a multi-channel fNIRS device, with long- and short-separation channels corresponding to the surgical skill montage shown in \autoref{fig:Montage}.}
    \label{fig:Surgical_example1}
\end{figure}

Surgical skill assessment data is simulated via 30-second trials with a stimulus start uniformly sampled from a range of  0 to 20 seconds, stimulus duration uniformly sampled from a range of 30-98 seconds, and a post-stimulus rest time uniformly sampled from a range of 0 to 30 seconds, which is consistent with the recorded trial start and end times present in the experimental data. The resulting recordings are subsequently cropped to 30 seconds for analysis, as described in \autoref{sec:experimental-data}.
All model hyperparameters used to simulate this dataset are listed in \autoref{tab:cortical_parameters} and \autoref{tab:surgical_parameters}. 
An example simulation of a surgical skill trial is shown in \autoref{fig:Surgical_example1}. 
In terms of computational efficiency, on an 11th Gen Intel Core i9-11900K, we simulate $500$ 30-98-second trials of surgical skill assessment data in under ten minutes for Jacobian-informed simulations and in one minute for Jacobian-free simulations. These simulation speeds do not include time for Jacobian generation. In comparison, the MNE-AR simulator takes $\approx90$ seconds to generate $500$ trials of surgical skill assessment data.

\subsection{MNE-AR Simulation Settings}
\review{To generate synthetic data using the MNE-AR model, we use MNE version 1.10 \citep{gramfort2013meg} and MNE-NIRS 0.7.1 \citep{luke2021analysis}. The simulation workflow generates the task response using nilearn's \texttt{hemodynamic\_models.glover\_hrf}, convolves it with an explicit stimulus boxcar, and normalizes the response to a unit peak. Baseline noise uses MNE's \texttt{add\_noise} with an IIR denominator. For MNE 1.1 compatibility, the code uses \texttt{compute\_whitener} for spatial coloring followed by scipy's \texttt{signal.lfilter}. Noise follows a shared AR(5) model across channels and wavelengths, with filter denominator: 1.0, -0.58853134, -0.29575669, -0.52246482, 0.38735476, 0.02428681.}

\review{Because confounding artifacts are common in the experimental data, automatic screening excluded invalid, flat, or extreme-variance/first-difference trials. For covariance estimation, experimental data were converted to $\Delta$OD and were mean-centered within each trial. The baseline measurement was taken to be the mean of the first five seconds of each trial. Within-trial scatter matrices were pooled, divided by the total degrees of freedom, and regularized using 5\% shrinkage toward the diagonal covariance. No denoising, temporal filtering, motion correction, or short-channel regression preceded covariance estimation.}

\end{document}